\documentclass[a4,twocolumn]{article}
\usepackage{natbib}
\usepackage{graphicx}
\usepackage{color} 
\usepackage{verbatim,framed}

\definecolor{orange}{rgb}{1.0,0.5,0.0}
\definecolor{aqgr}  {rgb}{0.0,1.0,0.6} 
\definecolor{viol}  {rgb}{0.8,0.6,1.0}

\title{Epigenetic Tracking: Towards a \\
       Project for an Artificial Biology}
       
\author{Alessandro Fontana$^{1}$ \\
\mbox{}\\
$^1$IEEE \\
alessandro.fontana@ieee.org}

\begin{document}
\maketitle

\begin{abstract}
This paper deals with a model of cellular growth called ``Epigenetic Tracking'', whose key features are: i) distinction bewteen ``normal'' and ``driver'' cells; ii) presence in driver cells of an epigenetic memory, that holds the position of the cell in the driver cell lineage tree and represents the source of differentiation during development. In the first part of the paper the model is proved able to generate arbitrary target shapes of unmatched size and variety by means of evo-devo techniques, thus being validated as a model of embryogenesis and cellular differentiation. In the second part of the paper it is shown how the model can produce artificial counterparts for some key aspects of multicellular biology, such as junk DNA, ageing and carcinogenesis. If individually each of these topics has been the subject of intense investigation and modelling effort, to our knowledge no single model or theory seeking to cover all of them under a unified framework has been put forward as yet: this work contains such a theory, which makes Epigenetic Tracking a potential basis for a project of Artificial Biology.
\end{abstract}

\section{Introduction}

This paper is concerned with a model of cellular growth called ``epigenetic tracking'' (described in \citep{AY08AX}), that belongs to the field or Artificial Embryology or Computational Development. The model is tested by its ability to generate arbitrary target shapes by means of evo-devo techniques, task which is taken as a measure of the model goodness and at which it appears to be quite successful. Subsequently, the implications of the model are explored, in relation to some key aspects of cell biology: embryogenesis, junk DNA, ageing and carcinogenesis: the model is shown able to produce artificial counterparts of each of these aspects (albeit with a reduced level of complexity). The paper is divided into two parts. The first part reviews the previous work in the field of artificial embryology (section 2.1), describes the model of cellular growth (section 2.2) and reports the results of the experiments performed (section 2.3). The second part shows how the model is able to generate artificial counterparts for each of the following key aspects of cell biology: embryogenesis (section 3), junk DNA (section 4), ageing (section 5) and carcinogenesis (section 6). Each of these sections opens with a description of the experimental evidence relevant to the aspect of biology considered, reviews the main existing models and theories and finally describes the artificial counterpart produced by epigenetic tracking. Scattered among more aspects, the topic of stem cells is also discussed. Finally, section 7 draws conclusions and outlines future research directions.

\section{The Model of Cellular Growth}

\subsection{Artificial Embryology related work}

The previous work in the field of Artificial Embryology (see \citep{AY03KB,AY03SM} for a comprehensive review) can be divided into two broad categories: the grammatical approach and the cell chemistry approach. The grammatical approach, originated by Lindenmayer \citep{AY68LX}, evolves sets of rules in the form of grammatical rewrite systems; the grammar can be context-free or context-sensitive and can utilise parameters; variations on this theme include using instruction trees or directed graphs in place of actual grammars. L-systems were employed as a means of describing the complex fractal patterns observed in nature and particularly the architecture of plants. The cell chemistry approach draws inspiration from the early work of Turing \citep{AY52TX}, who introduced a mathematical model of diffusion and reaction within a physical substrate. This approach attempts to mimic more closely how physical structures emerge in biology; cells are arranged in a physical space where simulated proteins can be sent as signals from one cell to another, as in nature.

Within the grammatical approach, Sims \citep{AY94SX} used directed graphs to evolve the body morphologies and neural networks of artificial creatures in a simulated 3D physical world; in these graphs, a node represents a body part and an edge specifies how body parts are connected. Using a domain similar to Sims', Hornby and Pollack \citep{AY02HP} applied L-systems to the simultaneous evolution of the body morphologies and neural networks of artificial creatures in a simulated 3D physical environment. Cangelosi, Nolfi and Parisi \citep{AY94CN} devised a model of neural development which includes cell division and cell migration in addition to axonal growth and branching; the development process shows successive phases of functional differentiation and specialisation. Gruau's Cellular Encoding \citep{AY96GW} uses grammar trees to encode steps in the development of a neural network starting from a single ancestor cell; the grammar tree contains developmental instructions at each node.

Within the cell chemistry approach, Random Boolean Networks (RBN's) were originally developed by Kaufmann as a model of genetic regulatory networks \citep{AY69KF}; in the context of the development of multi-cellular organisms, the attractors of RBN's are interpreted as the different ``cell types'' of the organism. De Garis \citep{AY99DG} developed a model for evolving shapes in 2D reproductive cellular automata; the model was successful in evolving convex shapes but non-convex shapes (e.g. the L-shape) presented a problem. Bongard and Pfeifer \citep{AY01BP} proposed a minimal model of ontogenetic development to evolve both the morphology and neural control of agents that perform a block-pushing task in a physically-realistic, virtual environment. Inspired by the cell adhesion process, Hogeweg \citep{AY03HX} developed a model to simulate morphogenetic processes such as cell migration or engulfing, achieving to evolve complex artificial organisms. Miller and Banzhaf \citep{AY03MB} developed artificial organisms (the french flag) based on a method called Cartesian Genetic Programming, which evolves a developmental program inside cells.

\subsection{The Model}

In our model the phenotype of the organism is represented as a 2-dimensional array of square-shaped cells, being each cell associated to a position on a grid. The development starts with a single cell placed in the middle of the grid, and unfolds in n artificial age steps, counted by the variable ``Global Organismal Age'' (GOA) that runs from 0 to n-1 (n is a parameter). The term ``global'' refers to the fact that the variable GOA is shared by all cells (and therefore it can be considered the global ``clock'' of the organism). To each cell four variables are associated:

\begin{itemize}
\item a flag indicating whether the cell is ``driver'' or ``normal''; 
\item the ``genome'', organised as an array of ``change operators'', which is identical in all cells;
\item the ``cell epigenetic type'' (CET), organised as an array of n integers (n is the number of artificial age steps), which is not identical in all cells; the CET is present only in driver cells; 
\item an integer representing the cell's colour. In the current implementation four colour values are foreseen (0,1,2,3), an extra value (-1) indicates cell absence. 
\end{itemize}

\begin{figure}[ht] \begin{center}
{\fboxrule=0.2mm\fboxsep=0mm\fbox{\includegraphics[width=8.01cm]{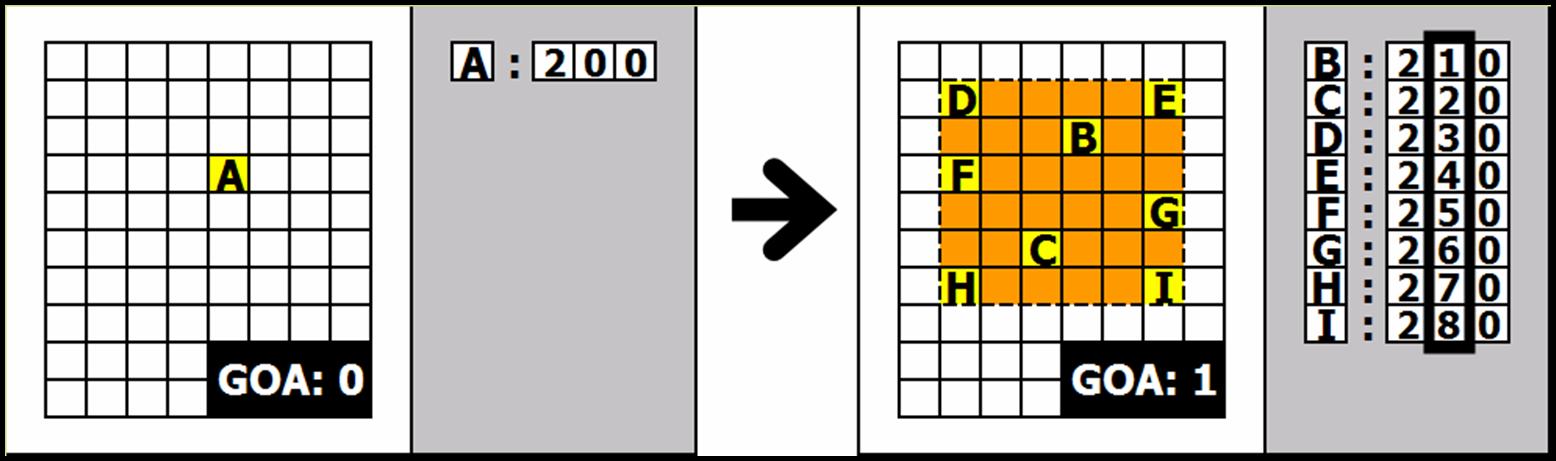}}} 
\caption{Example of proliferation; the column corresponding to the GOA value is highlighted with a thick frame (please note that column numbering starts from 0).}
\label{fig01}
\end{center} \end{figure}

Cells belong to two categories: ``driver'' cells (coloured in yellow in the figures) and ``normal'' cells (coloured in orange or blue). The basic difference between driver and normal cells is that the first can be instructed by the genome (by means of an operator whose left part matches the CET value of the cell) to proliferate or induce apoptosis in the surrounding area. Figure 1 shows an example: a driver cell associated to a CET value labelled with ``A''(called ``mother cell'') proliferates in an area around it (called ``change area'', delimited by a dotted line in the figure). While proliferating, it mostly generates normal cells (which fill the change area) and other driver cells, which are much fewer in number and ``dot'' the change area.

A key point is the assignment of the CET values on the newly created driver cells. To each new driver cell a new CET value is assigned, starting from the mother cell's CET value (the array [200] in the figure, labelled with ``A'') and adding 1 to the value of the i-th position of the array at each new assignment, where i is the current GOA value (1 in the figure, corresponding to the second column -column numbering starts from 0); with reference to the figure, the new driver cells are assigned the values [210],[220],... , labelled with ``B'',``C'', etc. In practise the variable CET holds the position of the driver cell in the {\bf driver cell lineage tree} or simply {\bf driver cell tree} (DCT, the set of all driver cells, having a tree structure): this ensures that the new CET values are all different from the mother's value and from each other. Whether one of these new CET values will become the centre of another proliferation event depends on the presence in the genome of an operator whose left part matches such value. 

The genome as we said is organised as an array of change operators (see figure 2). Excluding mutations, the genome is not modified during development and is identical in all cells. Each change operator has a left part and a right part. The left part consists of a variable called XOA, having the same structure of the variables GOA and a variable called XET, having the same structure of the variables CET: if the XOA value is equal to the GOA value and the XET value is equal to the CET value of a given driver cell, the operator (in this case we speak of a ``timed'' operator) is activated and the relevant code specified in the right part is executed for that cell; if the XOA value is -1, then the GOA value is ignored and the activation of the operator (in this case called ``non-timed'' operator) depends just upon the CET-XET match. The XET value is preceded by two parameters, one (OP, ``order of precedence'') indicating which operator takes precedence in case of multiple matches (XET values are not guaranteed to be unique) and one (ON) indicating whether the operator is ``structurally'' inactive or not. The right part of the operator has:

\begin{itemize}
\item a field with the coordinates of the rectangle which delimits the change area (row and column values of the north-west and south-east corners of the rectangle);
\item a field holding a ``master switch'' (MS0) that defines the shape of the change area (``rectangular'' -value = 0, ``diagonal left'' -value = 1, ``diagonal right'' -value = 2);
\item a field holding a second ``master switch'' (MS1) that defines the type of ``change event'' that is going to occur (``proliferation'' -value=0, ``apoptosis'' -value=1);
\item a field with a parameter (DT) that specifies the thickness of the diagonal (valid only if MS0=1 or 2);
\item a field with a parameter (CO) that specifies the colour of the newly created cells (both normal and driver).
\end{itemize}

In case of proliferation the change area is filled with newly created cells: most of the cells generated are normal cells, some are driver cells. The driver cells are much fewer in number (usually a ``linear normal to driver ratio'' of 5 has been used, corresponding to a 2-dimensional ratio of $5\cdot5=25$) and are deployed evenly on the change area (the precise algorithm to place the driver cells in not important as long as it ensures a uniform distribution). In case of apoptosis, all the cells contained in the change area ``die'', i.e. are deleted from the grid. The different types of change events (proliferation in different shapes -rectangular, diagonal left and right- and apoptosis) can be regarded as ``painting primitives'', i.e. basic painting actions that can be combined together to yield more complicated shapes. At the end of all change events the mother cell is always removed from the grid.

\begin{figure}[t!] \begin{center}
{\fboxrule=0.2mm\fboxsep=0mm\fbox{\includegraphics[width=8.01cm]{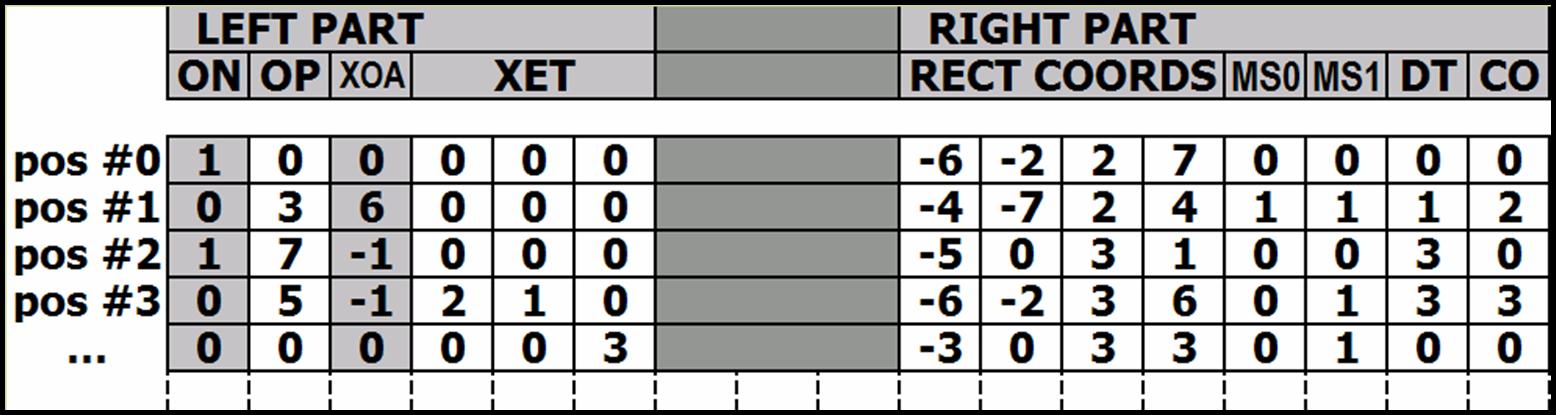}}} 
\caption{The genome.}
\label{fig02}
\end{center} \end{figure}

\begin{figure}[t!] \begin{center}
{\fboxrule=0.2mm\fboxsep=0mm\fbox{\includegraphics[width=8.01cm]{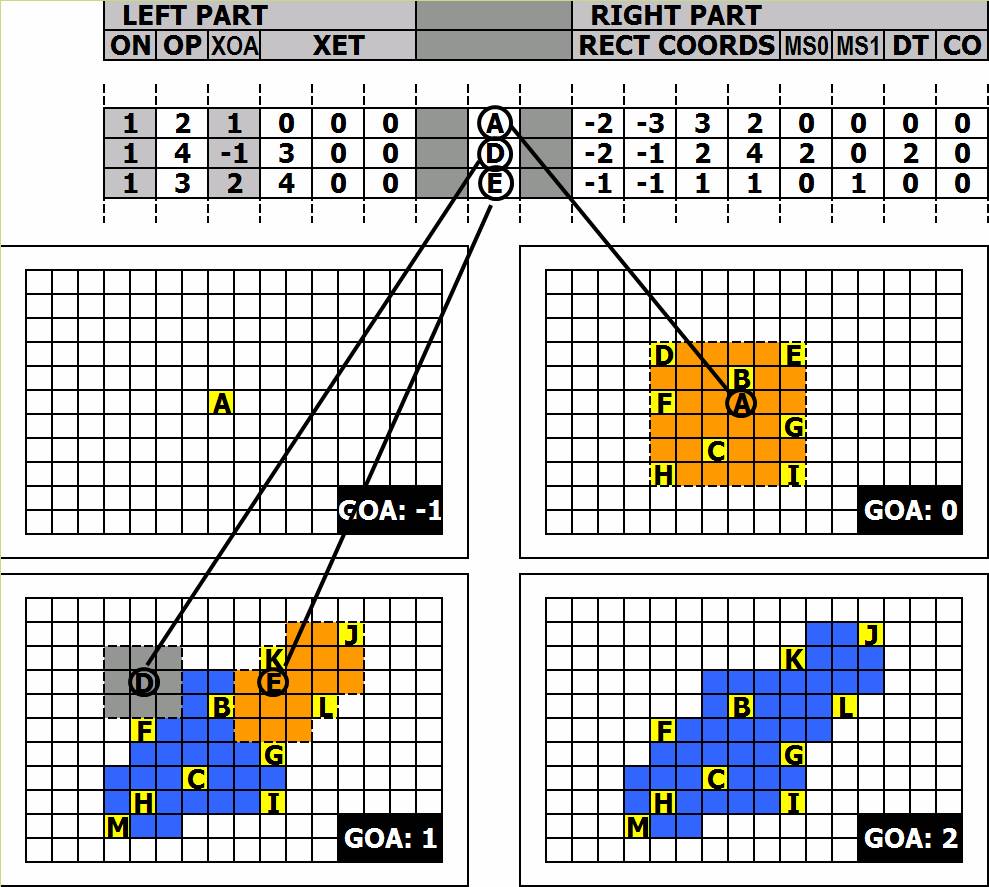}}} 
\caption{Example of development in four steps, driven by three operators. All ``snapshots'' are taken at the end of the relevant step; the zygote is present on the grid at the end of step -1, before the first age step, which is step 0. Normal newly generated cells are shown in orange, normal old cells in blue, driver cells in yellow.}
\label{fig03}
\end{center} \end{figure}

A special procedure is required if the change area is not empty. In this case the cells present must be either moved to other locations in the grid or removed altogether (overwritten). The solution chosen consists in first removing the cells present in the change area, carrying out the proliferation and finally redeploying the removed cells onto the grid. The order in which cells are removed and redeployed corresponds to their distance from the mother cell's position (experiments have been carried out with different types of distance).  

Figure 3 shows an example of development in four age steps (GOA=-1,0,1,2,) steered by three change operators, the first (a rectangular proliferation) triggered by the CET value labelled ``A'' in step 0, the second (a rectangular apoptosis) triggered by the CET value labelled ``D'' in step 1, the third (a diagonal right proliferation) triggered by the CET value labelled ``E'' also in step 1. The CET value ``A'' was present at the end of step -1 (starting point), The CET values ``D'' and ``E'' have been created in step 0. This example illustrates the ``core'' of the machine: a CET value produces a change event, which in turn produces other CET values, some of which produce other change events and so on, {\em in an indefinitely sustainable way}.

Let us summarise the key features of our model: i) the distinction bewteen ``normal'' and ``driver'' cells; ii) the implementation of the proliferation/apoptosis events in such a way that many cells are created/deleted at once; iii) the presence in driver cells of an epigenetic memory, that holds the position of the cell in the driver cell tree and represents the source of differentiation; iv) the mechanism of assignment of the CET values on the newly generated driver cells during a proliferation event, which ensures that each new driver cell is assigned a new, previously unseen CET value. We argue that this set of features is unique to our model and allows to reach a high level of performance in terms of both size and variety of the evolved shapes, as it will be shown in the next section. 
 
\subsection{Experiments}

\begin{figure}[t!] \begin{center}
{\fboxrule=0.2mm\fboxsep=0mm\fbox{\includegraphics[width=8.01cm]{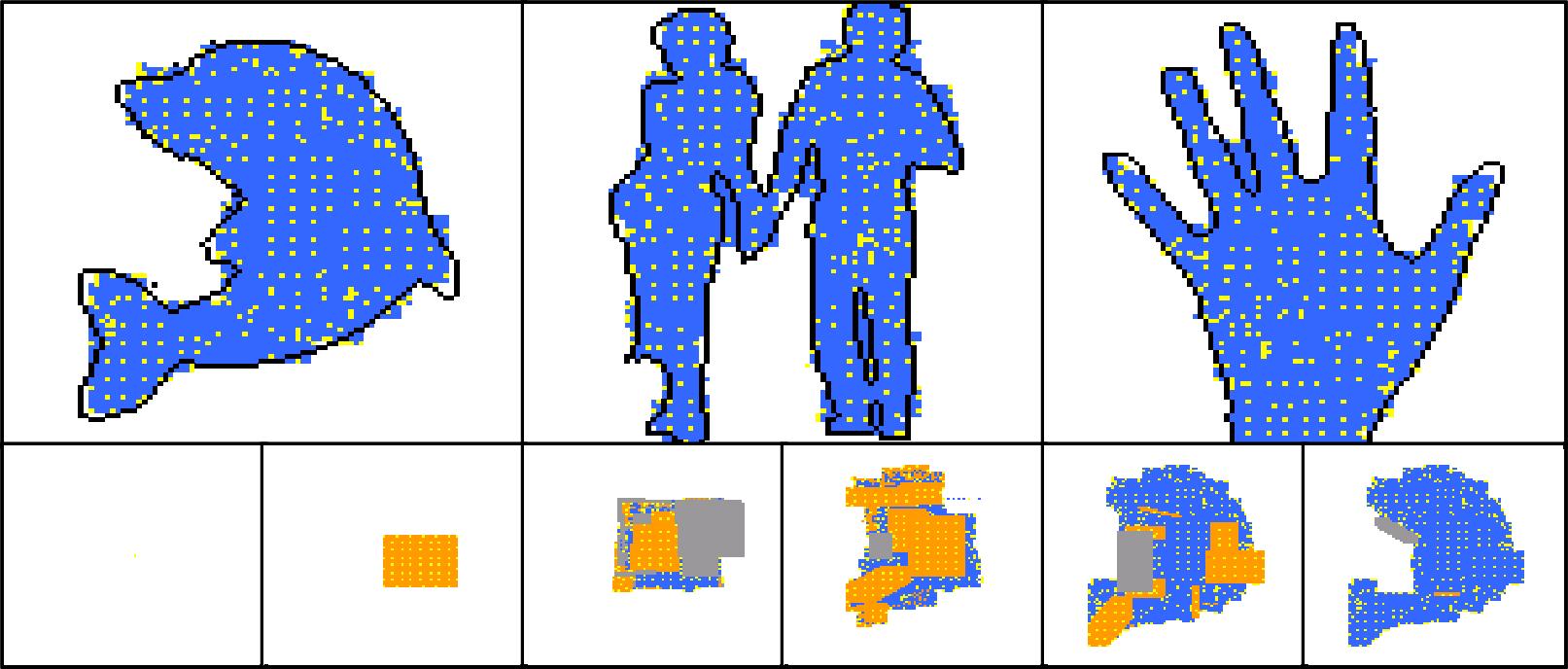}}} 
\caption{The dolphin, the couple and the hand (dynamical view, target contour superimposed). In the lower part the development sequence of the dolphin.}
\label{fig04}
\end{center} \end{figure}

The model described in the previous section has been tested on the problem of artificial morphogenesis and cell differentiation achieved by means of evolutionary techniques, i.e the task of generating predefined 2-dimensional shapes by evolving genomes that guide the development of the shape starting from a single cell. The experimental procedure consists in evolving a population of genomes, at each generation letting the development unfold for each genome (starting from a single cell with CET = [0,...,0] placed in the middle of the grid and running GOA from 0 up to a maximum value), and then using the adherence of the shape at the end of the development to the target shape as fitness measure. The genetic population is composed of 600 individuals (represented as strings of quaternary digits), undergoing elitism selection for up to 20000 generations. The parameters of the Genetic Algorithm (GA) are 50\% single point crossover, mutation rate of 0.1\% per digit. The fitness function formula is the same adopted by H. de Garis \citep{AY99DG}:
\begin{eqnarray}
\mathrm{F}=(\mathrm{ins}-\mathrm{outs})/\mathrm{des}
\end{eqnarray}
where ins is the number of cells of the evolved shape falling inside (and matching the colour of) the target shape, outs is the number of cells of the evolved shape falling outside the target shape, des is total number of cells of the target shape. At each age step two views are possible for the developing shape: one that shows the ``dynamics'' of the change operators, and one that shows the colours of cells. In the ``dynamical view'' driver cells are coloured in yellow, normal cells are coloured in orange if they have been just (i.e. in the current step) created, in blue if they have been created in one of the previous steps; areas where cells have been deleted by an apoptosis event are coloured in grey. In the ``colour view'' (which of course makes sense only for colour targets) cells are shown with their actual colours. 

\begin{figure}[t!] \begin{center}
{\fboxrule=0.2mm\fboxsep=0mm\fbox{\includegraphics[width=8.01cm]{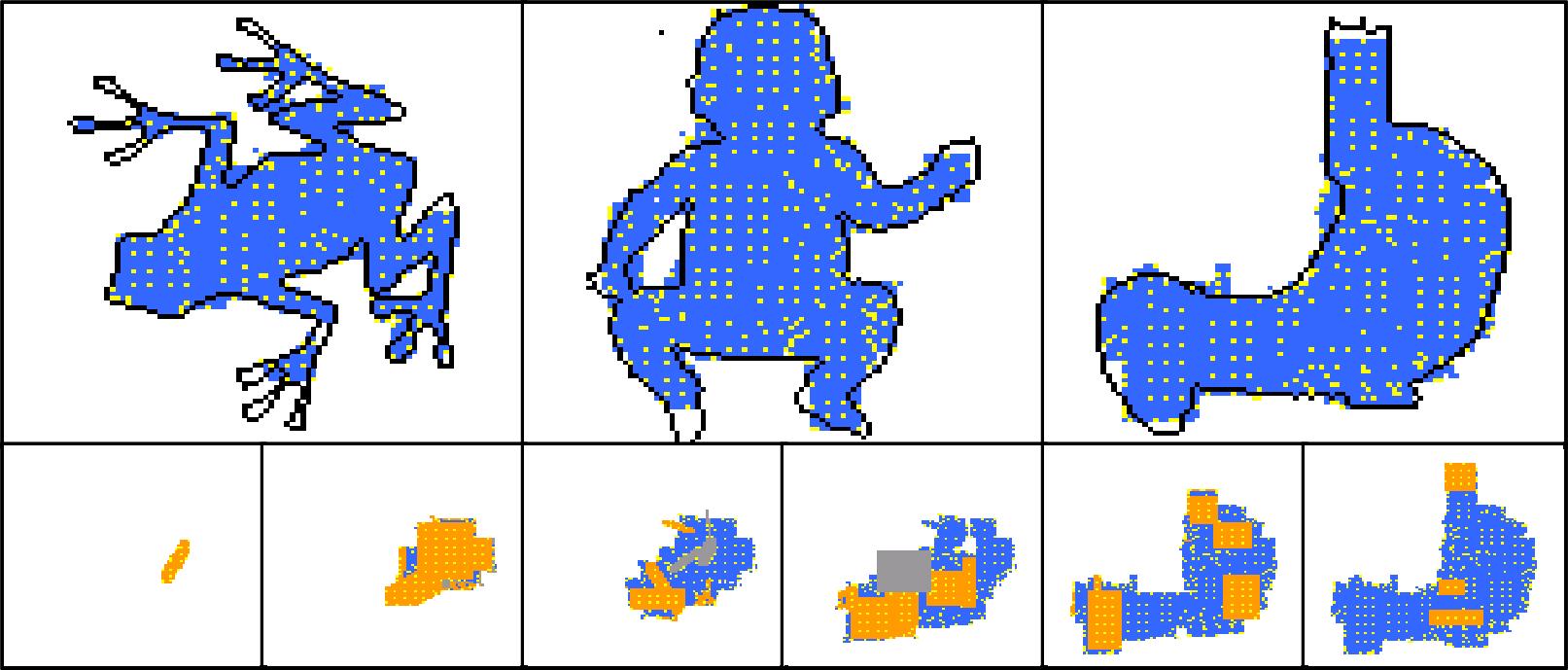}}} 
\caption{The frog, the baby and the stomach (dynamical view, target contour superimposed). In the lower part the development sequence of the stomach.}
\label{fig05}
\end{center} \end{figure}

Simulations have been conducted with a number of different target shapes; the targets have been chosen with the objective of testing the method on shapes as diverse as possible, to prove its effectiveness in generating any kind of shape. All targets are 100x100 multi-cellular arrays: the limited computational resources available prevented us from putting to a test larger shapes. Figures 4-7 show the results of simulations conducted with black-and-white and colour targets (in some simulations slightly different painting primitives have been used). As we can see, all target shapes have been approximated to a good degree; colour targets have proved more difficult to evolve, as one may have expected. To our knowledge, no other method is able, by means of evo-devo techniques, to generate target shapes with this size and variety. The remainder of this paper will be dedicated to showing how the model, beside being successful at the task it was designed for, relates to some key aspects of biology, namely embryogenesis, junk DNA, ageing and carcinogenesis. Embryogenesis will be examined first.
 
\section{Embryogenesis}

\subsection{Experimental evidence on embryogenesis}

Embryogenesis is the process by which the embryo is formed and develops. It starts with the fertilisation of the egg, then called a zygote. The zygote undergoes rapid cell divisions with no significant growth, producing a cluster of cells that is the same size as the original zygote, called {\bf morula}. The next stage is the {\bf blastula}, a spherical layer of cells surrounding a fluid-filled or yolk-filled cavity; mammals form then a structure called {\bf blastocyst}, characterised by an inner cell mass not present in the blastula. During next stage, called {\bf gastrulation}, cells migrate to the interior of the blastula, forming three (in triploblastic animals) germ layers, referred to as ectoderm, mesoderm and endoderm. At some point after the different germ layers are defined, {\bf organogenesis} begins; the first stage in vertebrates is called neurulation, where the neural plate folds forming the neural tube, but from now on embryogenesis follows no common pattern among the different taxa of the animal kingdom.

\begin{figure}[t!] \begin{center}
{\fboxrule=0.2mm\fboxsep=0mm\fbox{\includegraphics[width=8.01cm]{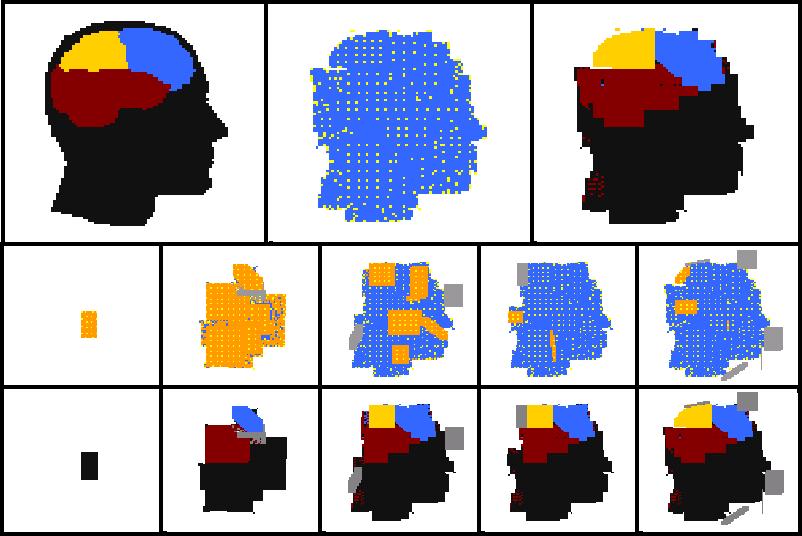}}} 
\caption{The head. In the upper part, on the left the target shape, in the middle the best evolved shape in dynamical view and on the right the best evolved shape in colour view. In the lower part some development steps.}
\label{fig06}
\end{center} \end{figure}

Embryogenesis is the result of three coordinated processes: morphogenesis, cell growth and cell differentiation. Morphogenesis is concerned with the particular aspect of embryogenesis relevant to the shapes of tissues, organs and entire organisms and the positions of the various specialised cell types. Cell growth and differentiation can take place in cell culture or inside of tumor cell masses without the normal morphogenesis that is seen in an intact organism. In the human embryo, the change from a cluster of nearly identical cells at the blastula stage to a post-gastrulation embryo with structured tissues and organs is controlled by the genetic ``program'' and can be modified by environmental factors. A key role in the process of embryogenesis is thought to be played by stem cells.

Stem cells are cells found in most, if not all, multi-cellular organisms. The classical definition of a stem cell requires that it possess two properties: i) {\bf self-renewal} (the ability to go through numerous cycles of cell division while maintaining the undifferentiated state) and ii) {\bf potency} (the capacity to differentiate into specialised cell types). Two types stem cells exist: embryonic stem cells (found in the inner cell mass of the blastocyst) and adult stem cells (found in adult tissues). {\bf Embryonic stem (ES) cells} are totipotent: this means they are able to differentiate into all derivatives of the three primary germ layers, including each of the more than 220 cell types in the adult body; when given no stimuli for differentiation, embryonic stem cells maintain totipotency through multiple cell divisions. {\bf Adult stem cells} are undifferentiated cells found throughout the body after embryonic development that divide, to replenish dying cells and regenerate damaged tissues; pluripotency distinguishes adult stem cells cells from totipotent embryonic stem cells: they are only able to form a limited number of cell types. Examples of adult stem cells are hematopoietic stem cells and colon stem cells. 

\begin{figure}[t!] \begin{center}
{\fboxrule=0.2mm\fboxsep=0mm\fbox{\includegraphics[width=8.01cm]{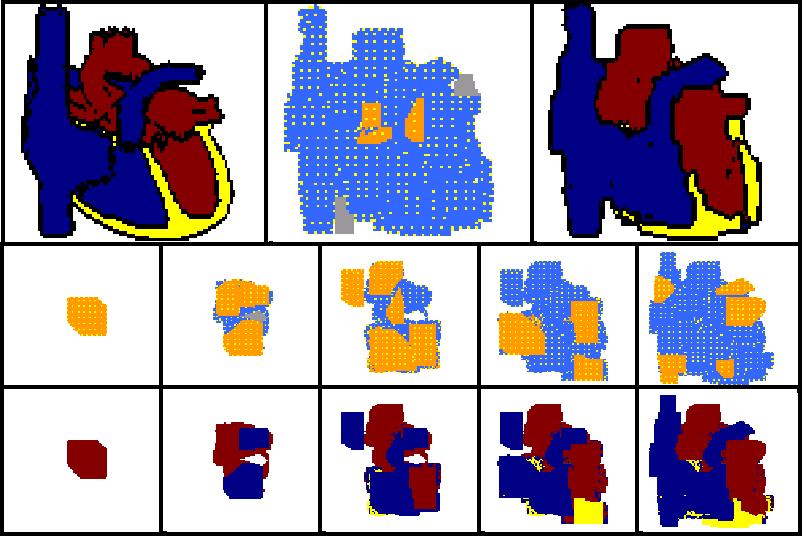}}} 
\caption{The heart. In the upper part, on the left the target shape, in the middle the best evolved shape in dynamical view and on the right the best evolved shape in colour view. In the lower part some development steps.}
\label{fig07}
\end{center} \end{figure}

\subsection{Artificial embryogenesis}

The model of cellular growth called ``epigenetic tracking'' has been tested experimentally with the problem of artificial morphogenesis and cell differentiation, implemented respectively through the shaping and colouring of cellular sets: therefore its interpretation as a model of artificial morphogenesis and cell differentiation is straightforward. Given a model of cellular growth, a key question is how to measure the model's goodness. It is our belief that a very important ingredient in assessing a model's goodness is represented by its evolvability: in this respect our model's ability to evolve complex target shapes can be considered experimentally demonstrated. The level of complexity of the shapes generated is of course still very far from the level of complexity displayed by nature, but nevertheless very high compared to other growth models.  

In the model of cellular growth proposed, driver cells take the role of embryonic stem cells. The artificial counterpart of the property of ``potency'' (characteristic of ES cells), is represented by the capacity of driver cells to give rise to cells of any colour. As far as the ``self-renewal'' property is concerned, the situation is a bit different, in that ES cells, when proliferating, give rise to other identical ES cells, while in our model this is not strictly true: driver cells, when proliferating, give rise to other driver cells which have different CET values. The CET value of the mother and those of the daughters is of course similar (they share the upper sub-tree), but not identical. Driver cells have also one property in common with Spemann's organisers: if a driver cell (a Spemann's organiser) destined to give rise to a certain shape (embryo) part is moved to a different position of the growing shape (embryo), that shape (embryo) part will grow in the new, ectopic position.

The artificial genome corresponds to the natural genome: it is defined as that part of genetic information which is shared by all cells. The cell epigenetic type (CET) corresponds to the cell epigenetic memory (stored in DNA's methylation patterns); it represents the part of genetic information that is different from cell to cell and, as such, it constitutes the primary source of the information necessary for cellular differentiation. The CET can be thought as the provider of the artificial counterpart of the ``first transcription factor'', that gives origin to the whole cascade of gene activations that build-up the transcriptome and determine the cellular identity; for driver cells such factor comes from within, while for non-driver cells it has to be supplied from the outside.  

Speaking about features of the model that lack biological plausibility, at present inter-cellular signals are not modelled; as a result, the local environment has no influence on cell fate determination, which is in contrast with the biological evidence. Moreover, in our model any change event is carried out by a single operator activated by a single CET value; in other words, we have a genetic regulatory network in which the outputs are connected directly to the inputs, with no ``hidden layer''; we know by contrast that in biological systems events are determined by the interplay of many genes. It is in our plans to add these features to the model and it is our opinion that such additions will be beneficial in terms of robustness to perturbations, with no negative impacts on evolvability. 

We end this section by noting that, even if the target shape would be coincident with the picture of a real living being (an animal, for instance), the sequence of morphogenetic steps our algorithm would follow to develop the shape would in general be different from the actual sequence followed by the animal embryo during its development. This brings us to conclude that our algorithm is {\em not} able to reproduce exactly biological embryogenesis; rather, it is able to produce a phenomenon that can be considered of analogous nature. Needless to say, biological reality is much more complex compared to our experiments, which are nonetheless significantly more complex, in terms of size and variety of the shapes generated, than most (if not all) related work.     

\section{Junk DNA}

\subsection{Experimental evidence and theories on junk DNA}

In molecular biology, ``junk DNA'', or ``noncoding DNA'', is a collective label for the portions of the DNA sequence of a genome for which no function has been identified. About 95\% of the human genome has been designated as ``junk'', including most sequences within introns and most intergenic DNA. While much of this sequence may be an evolutionary artifact that serves no present-day purpose, some junk DNA may function in ways that are not currently understood. Moreover, the conservation of some junk DNA over many millions of years of evolution may imply an essential function: according to a comparative study of over 300 prokaryotic and over 30 eukaryotic genomes \citep{CY08AX}, eukaryotes appear to require a minimum amount of non-coding DNA (in humans the predicted minimum is about 5\% of the total genome). Some chromosomal regions are composed of the now-defunct remains of ancient genes, known as {\bf pseudogenes}, which were once functional copies of genes but have since lost their protein-coding ability. As much as 25\% of the human genome is recognisably formed of {\bf retrotransposons} \citep{CY02DB}: new research suggests that genome size variation in at least two kinds of plants is mostly because of retrotransposons \citep{CY06PX}.

There are some hypotheses, none conclusively established, for how junk DNA arose and why it persists in the genome:

\begin{itemize}
\item Junk DNA might provide a reservoir of sequences from which potentially advantageous new genes can emerge. In this way, it may be an important genetic basis for evolution;
\item Some junk DNA could be spacer material that allows enzyme complexes to form around functional elements more easily. In this way, it could serve a useful function regardless of the actual base sequence;
\item Some portions of junk DNA could serve presently unknown regulatory functions, controlling the expression of certain genes during the development of an organism from embryo to adult \citep{CY05WX}.
\end{itemize}

\subsection{Artificial junk DNA}

In our model, at any given moment in the course of evolution, an individual's driver cell tree generated during development can be divided into i) driver cells (i.e. their CET values) that activate an operator during development and ii) driver cells that {\em do not} activate any operator during development. In the same way the individual's genome is composed by i) operators (i.e. their XET values) that become active during development and by ii) operators that {\em do not} become active during development. By analogy with real genomes, elements in the two categories labelled with ii) can be defined as ``junk'' driver cells (CET values) and ``junk'' operators (XET values) respectively. A schematic representation of this distinction is given in figures 8 and 9: figure 8 shows an example of development in four steps and figure 9 shows the corresponding driver cell tree and genome, where grey circles represent junk elements. For ease of reference the set of driver cells active during development will be called DCT-D (``Driver Cell Tree D'', D for ``development''), while the set of driver cells not active during development will be called DCT-I (I for ``inactive''). Analogously, the set of operators active during development will be called GEN-D (``Genome D'') and the set of operators not active during development will be called GEN-I. In the rest of this section we will argue that the presence of junk in both the driver cell tree and the genome is an inescapable phenomenon, intimately linked to the functioning of the epigenetic tracking machine.

\begin{figure}[t!] \begin{center}
{\fboxrule=0.2mm\fboxsep=0mm\fbox{\includegraphics[height=5.4cm]{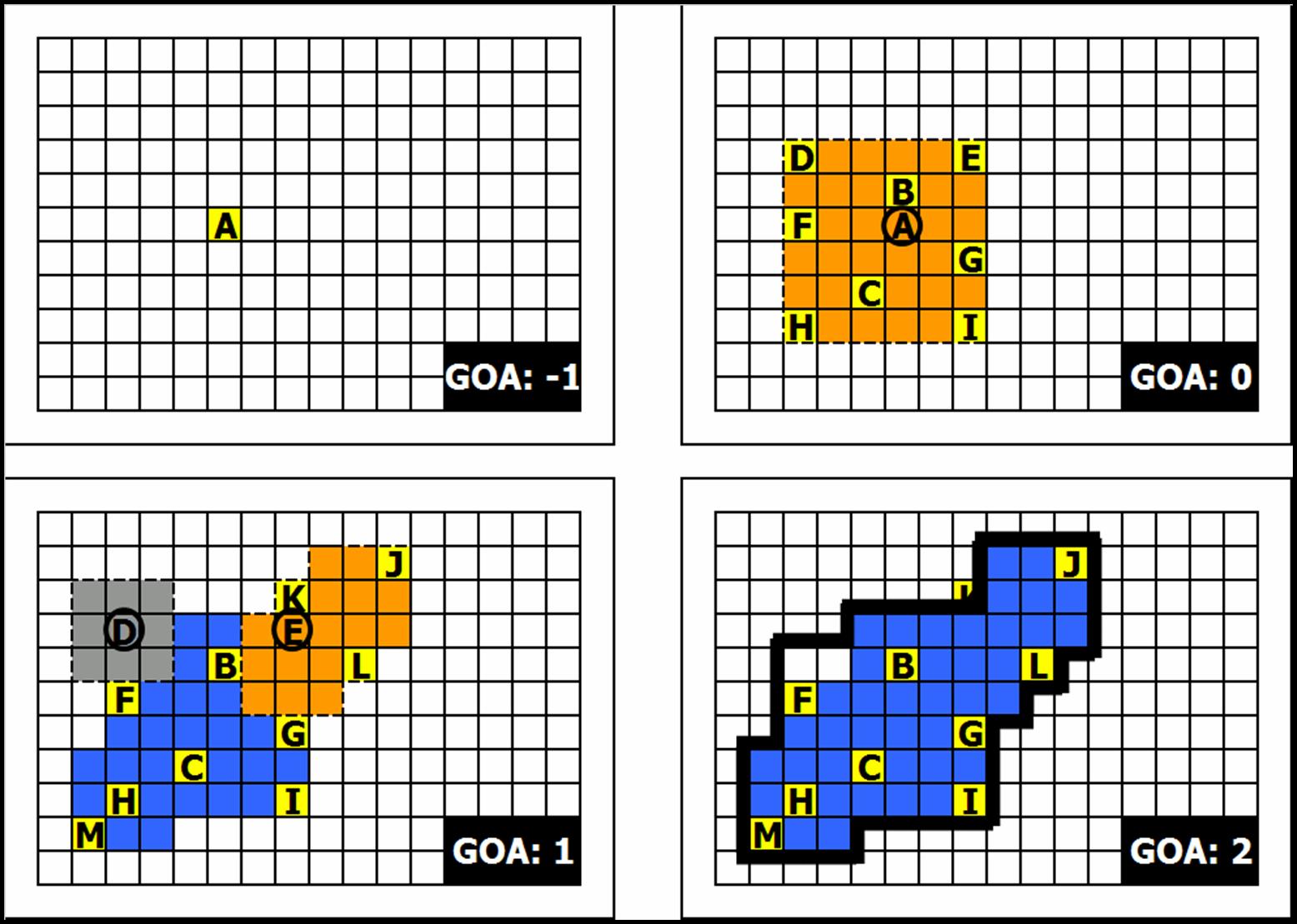}}} 
\caption{Example of development in four age steps steered by three change operators (target shape superimposed in GOA=2).}
\label{fig08}
\end{center} \end{figure}

The variable CET, as described in section 2, is organised as an array of n integers, where n is the number of artificial age steps. A legitimate question is which considerations motivated this design choice and, in particular, why an array structure was preferred to a simpler scalar structure. In reality the first idea was to use a scalar CET value, along with a global counter, subject to the following rules:

\begin{itemize}
\item The global counter CNT stores the last CET value assigned in a proliferation event occurred anywhere in the shape; it is incremented by one at each new assignment;
\item The zygote's CET value is zero. Each time a proliferation event takes place, the CET value assigned to the first newly created driver cell is the value held by the global counter; subsequent values are determined adding one at each new assignment (the global counter is updated correspondingly).
\end{itemize}

\begin{figure}[t!] \begin{center}
{\fboxrule=0.2mm\fboxsep=0mm\fbox{\includegraphics[height=5.4cm]{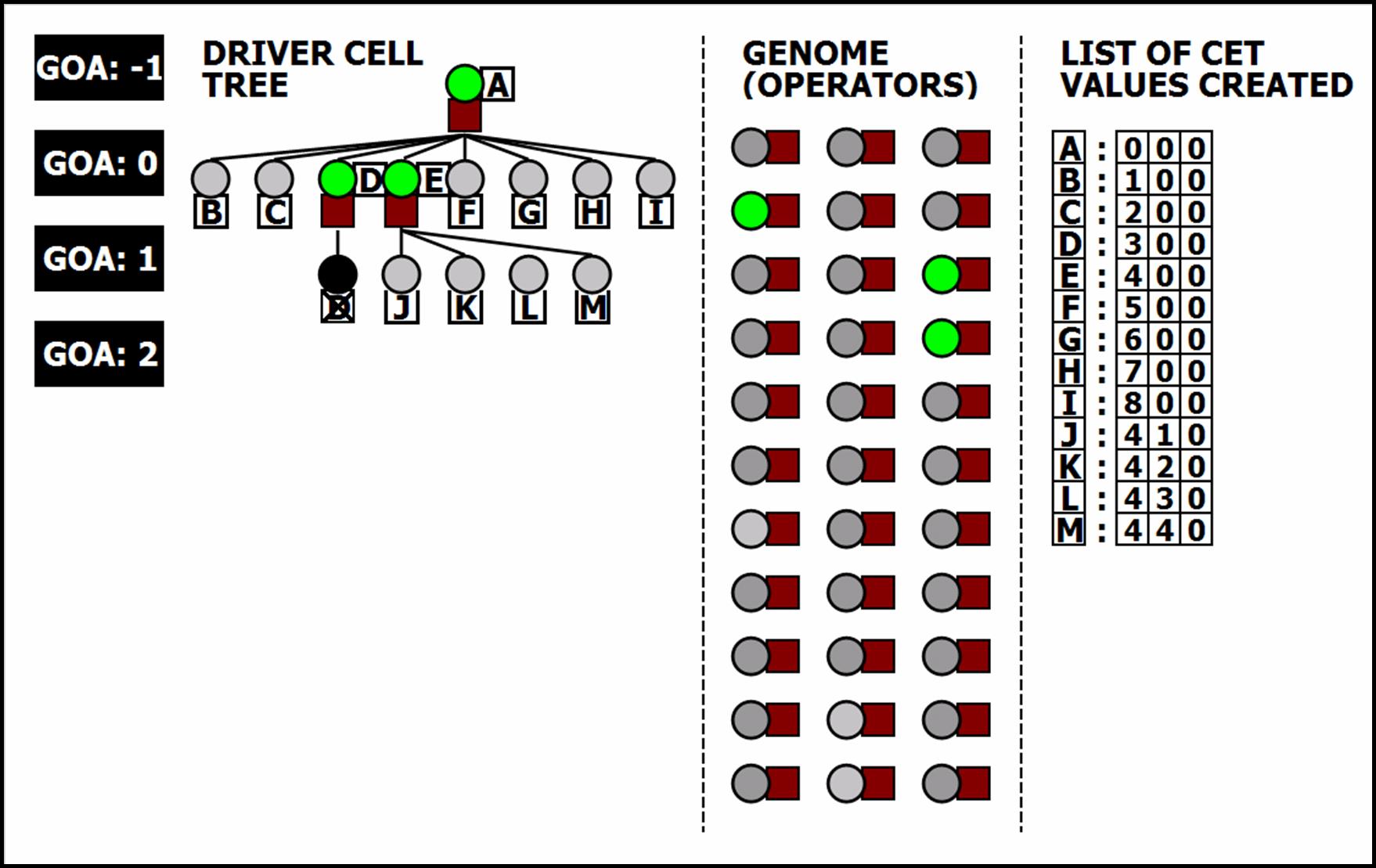}}} 
\caption{The corresponding driver cell tree. Grey circles represent CET values that match no XET values and XET values that match no CET values (junk elements); green circles represent CET values and XET values that match; brown squares represent right parts of operators.}
\label{fig09}
\end{center} \end{figure}

This ``scalar approach'', an example of which is given in figure 10, has two major drawbacks. The first drawback is due to the need of a global variable (the global counter), that has to be accessed constantly by all organism's cells, in order to be kept updated: for this reason such approach is {\em not} biologically plausible. The second drawback is that it creates ``dependencies'' between parts of the developing shape that would otherwise be unrelated. Figure 11 shows an example of such a dependency. Panel A represents the starting point of development, with GOA=-1 and CNT=1. Let's now imagine that evolution invents the developmental path A-B-C (``path X''), in which the part of the shape circled in red remains undeveloped. The most straightforward way evolution has to develop such part is to cast a proliferation event on the driver cell with the CET value 4 (circled in black), thus exploring a new developmental path (``path Y'').

In this new path, such proliferation event takes place in GOA=1 (panel D). As a result, the drivers with CET values 6-9 are either not existing any more or are placed in different positions (compared to path X). As a further result, the operators acting on such values have now random effects and the area circled in panel D remains unfilled, bringing the fitness of path the Y individual down compared to the fitness of the path X individual. Faced with choosing between paths X and Y, the Genetic Algorithm will therefore choose path X, which has an overall higher fitness (the unfilled area in panel C is smaller than the unfilled area in panel D): the ultimate effect is that path Y will not be selected for and the circled region in panel C will remain unfilled. The Genetic Algorithm has still the possibility to use a timed operator (i.e. an operator whose activation is conditioned to GOA taking a certain value -in this case greater than 2), but such operators are harder to evolve and the effectiveness of the evolutionary process is in any case reduced.

\begin{figure}[t!] \begin{center}
{\fboxrule=0.2mm\fboxsep=0mm\fbox{\includegraphics[height=5.4cm]{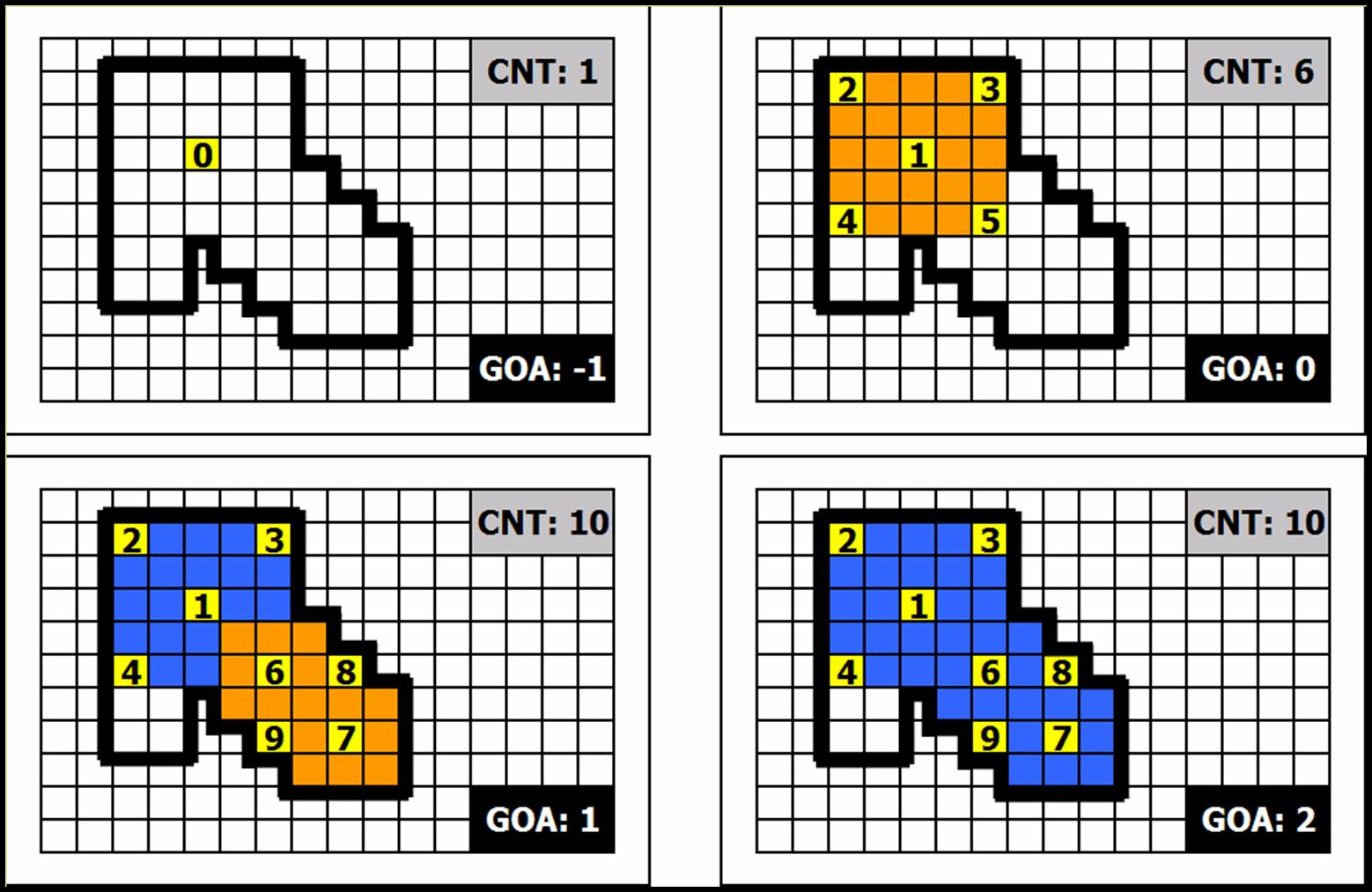}}} 
\caption{Example of development in four steps carried out with a scalar CET and a global counter.}
\label{fig10}
\end{center} \end{figure}

The introduction of an array version for the CET overcomes these two problems at once. Besides being biologically plausible, it decouples the CET values generated in different proliferation events, thus eliminating the risk of unwanted dependencies. Unfortunately a new problem, related to the size of the Genetic Algorithm's search space, arises. With an array of n numbers, each with, say, 10 possible values, the total number of array values is $10^{n}$: this is the size of the space the GA has to  search. As the value of n grows beyond values not too small, this number quickly becomes very large and the GA search space becomes unmanageable, virtually bringing evolution to a halt. The solution to this problem consists in a procedure called ``Germline Penetration'' (see figure 12). Such procedure acts on the genome of each individual at the end of development, copying at random the CET values occurred during development onto the XET values of change operators in the genome, as a suggestion for the GA for where to search: with this procedure in place, the effectiveness of the GA is restored. To avoid disrupting development, the ``transplanted'' operators are set as inactive, so that they start their ``career'' as junk operators.  

\begin{figure}[t!] \begin{center}
{\fboxrule=0.2mm\fboxsep=0mm\fbox{\includegraphics[height=5.4cm]{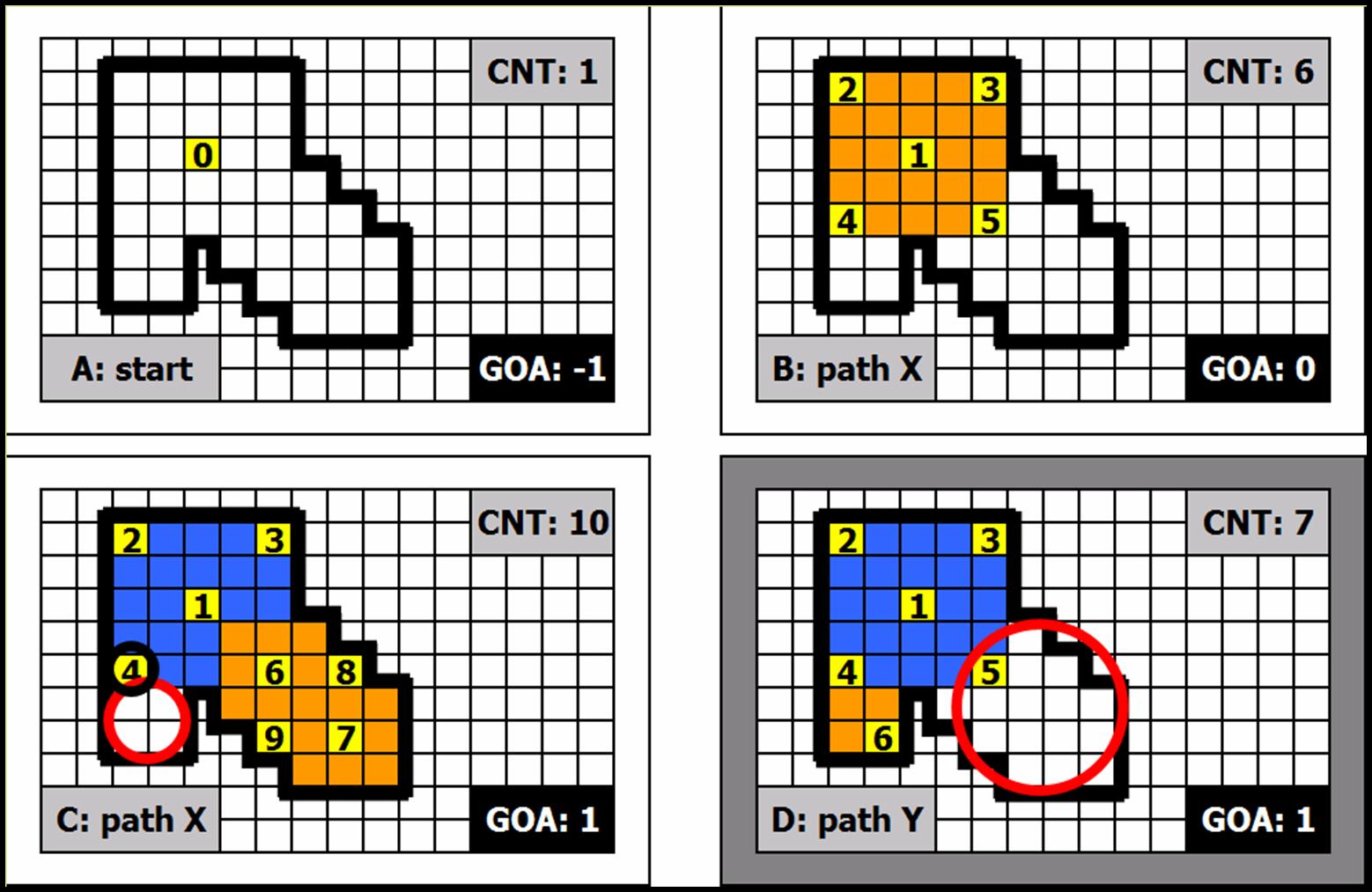}}} 
\caption{The utilisation of a scalar CET limits the effectiveness of the GA in exploring the search space.}
\label{fig11}
\end{center} \end{figure}

\begin{figure}[t!] \begin{center}
{\fboxrule=0.2mm\fboxsep=0mm\fbox{\includegraphics[height=5.4cm]{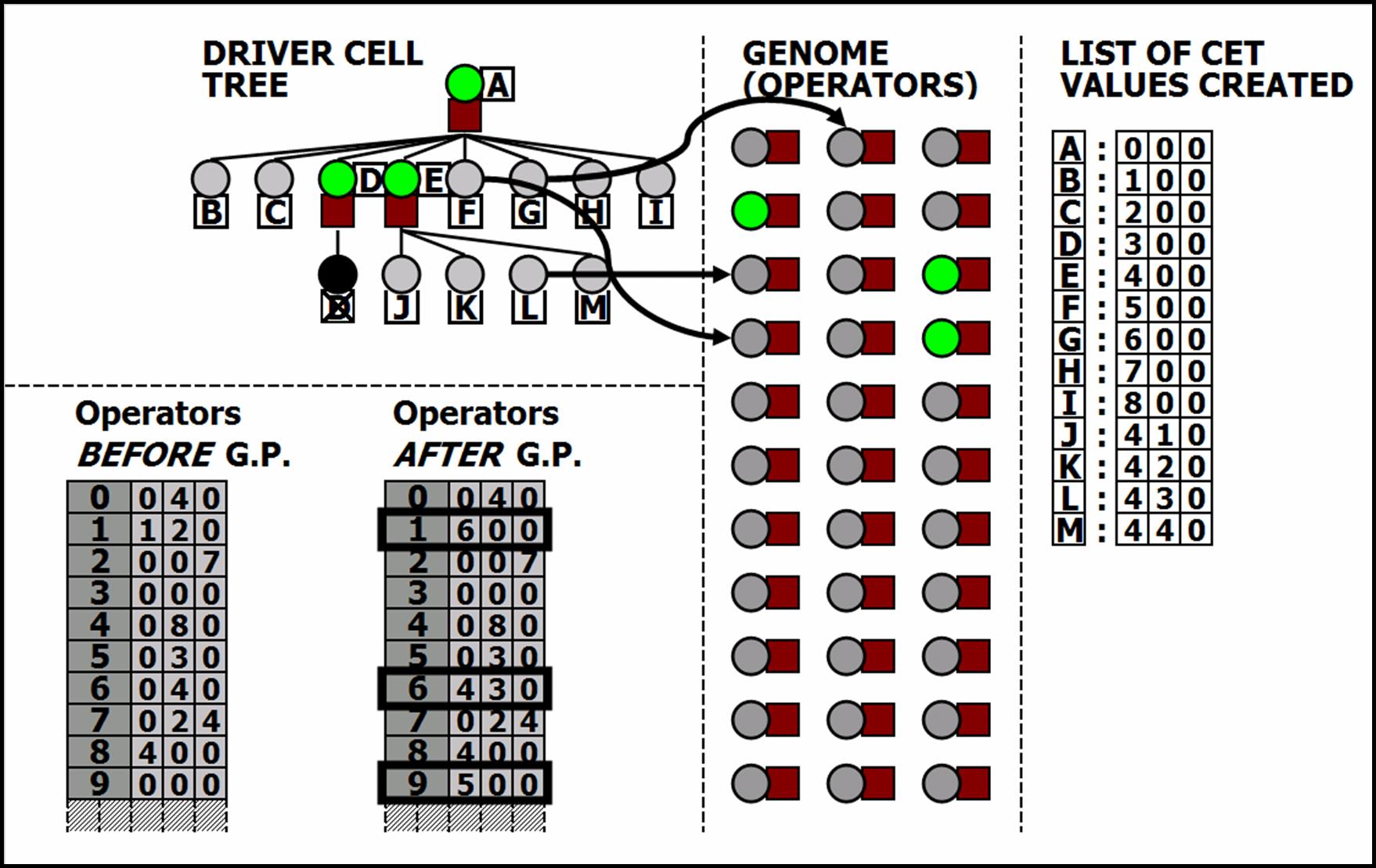}}} 
\caption{Germline Penetration transfers values from the driver cell tree to the genome (copied values are highlighted with a thicker border).}
\label{fontana12_4xglobcnt2}
\end{center} \end{figure}

With the previous considerations in mind we can now turn our attention to junk DNA. The epigenetic tracking machine, the way it is conceived, cannot do its job without generating a lot of junk CET values. In the development of the dolphin (figure 4), for instance, 471 CET values are generated and only 27 are used; in the development of the frog (figure 5), 555 CET values are generated and only 25 are used; on average, in the experiments performed, the average ratio (CET values used / CET values generated) is around 5\%. The presence of such a high percentage of unused CET values seems to be a waste of resources: we could ask if there is a way to reduce it, improving the perceived efficiency of the technique. The most straightforward way to reduce the amount of unused CET values consists in decreasing the ratio between driver and normal cells in proliferation events: for instance, instead of the value of 1:25 used in our experiments, we could use a value of 1:10. This would cause the evolved shapes to have, at any age, a sparser distribution of driver cells (a lower density of ``yellow dots'') and, as a result, the ``paintbrush'' would become less precise and evolution of development would become harder.

So, there seems to be a trade-off between the precision of painting and the density of driver cells (and hence the percentage of junk CET values): the second aspect cannot be improved without worsening the first. Since the effectiveness in evolving shapes is the primary objective of the method, in this regard we are not prepared to make concessions: therefore, a certain amount of unused CET values must be reckoned with. The conclusion of the discussion so far can be expressed by saying that the following two facts have emerged as inescapable, inherent characteristics of our method:

\begin{itemize}
\item the presence of a high percentage of junk CET values, to allow for a sufficient precision of the painting process;
\item the need to use the procedure called Germline Penetration, to allow the GA to work with an array-structured CET variable, thus eliminating the drawbacks deriving from the use of a scalar CET variable.
\end{itemize}

Putting these two elements together, it would not be surprising to observe that Germline Penetration acts like a shuttle, transferring junk CET values from the driver cell tree onto a corresponding number of junk XET values in the genome (with the hope that they meet each other and will not be junk anymore!): actually, this is exactly what we saw happening in our experiments. Since in such experiments the total number of change operators is kept fixed, the percentage of junk XET values in the genome does not, in general, match the percentage of junk CET values in the driver cell tree. Going back to the frog example, since the number of active operators is 25 and the total number of operators in the genome is fixed at, say, 100, percentage of junk XET values is (100-25)/100 = 75\%, significantly lower than the corresponding percentage of junk CET values (95\%): should we allow the genome to have a flexible size, we could expect these two percentages to be roughly equal.

The conclusion of this section, therefore, is that the presence of junk information in both the driver cell tree and the genome is a fact that is inescapably connected to the core of the epigenetic tracking machine, a requirement essential to its {\em evolvability}. Finally, we like to conclude with two observations of a more speculative nature. Firstly, should we hypothesise the existence of a mechanism akin to Germline Penetration also in biological systems, we would be naturally led to think of mobile DNA elements, or transposons, as the actual device used to carry the CET values from the biological equivalent of driver cells, spread throughout the body, to the germline cells, where they would deliver the recipe of current development as a suggestion for future improvements. Secondly, we note that, by allowing the developmental history of an organism to influence its genome and therefore to be passed on to the next generation, the mechanism of Germline Penetration adds a Lamarckian touch to the Darwinian evolution implemented by the Genetic Algorithm. 

\section{Ageing}

\subsection{Theories on ageing}

Ageing is the accumulation of changes in an organism over time, leading to a steady decline in bodily functions. It is not a universal phenomenon: in a few simple species, ageing is negligible and cannot be detected (such species are not immortal, however, as their members will eventually fall prey to trauma or disease). The only measure that was proved to be effective in slowing the ageing process in a wide variety of species is caloric restriction. Theories that explain ageing have generally been divided between stochastic and programmed theories of ageing. Stochastic theories blame environmental impacts that induce cumulative damage on living organisms at various levels as the cause of ageing. Programmed theories imply that ageing is regulated by biological clocks operating throughout the life span; this regulation would depend on changes in gene expression that affect the systems responsible for maintenance, repair and defense responses.

Within the first category, the {\bf wear-and-tear theory} maintains that changes associated with ageing are the result of chance damage that accumulates over time (not unlike the ``ageing'' of a mechanical device). According to the {\bf somatic mutation theory}, ageing results from damage to the genetic integrity of the body's cells. The {\bf accumulative-waste theory} points to a buildup of cells of waste products that presumably interferes with metabolism.
The {\bf free-radical theory} is based on the idea that free radicals create damage that gives rise to symptoms we recognise as ageing.

Within the second category, the {\bf ageing-clock theory} argues that ageing results from a preprogrammed sequence, as in a clock, built into the operation of the nervous or endocrine system of the body. In rapidly dividing cells the shortening of the telomeres (structures at the ends of chromosomes that have experimentally been shown to shorten with each successive cell division) would provide just such a clock. The {\bf reproductive-cell cycle theory} is built around the idea that ageing is regulated by reproductive hormones that act in an antagonistic pleiotropic manner via cell cycle signaling: they promote growth and development early in life in order to achieve reproduction, but later in life become dysregulated and drive senescence. {\bf Evolutionary theories} share the main assumption that ageing has evolved because of the increasingly smaller probability of an organism still being alive at older age, due to predation and accidents. It is thought that strategies which result in a higher reproductive rate at a young age, but shorter overall lifespan, result in a higher lifetime reproductive success and are therefore favoured by natural selection. The {\bf disposable soma theory} \citep{DY77KX} and the {\bf mutation accumulation theory} belong to this group.

\begin{figure}[t!] \begin{center}
{\fboxrule=0.2mm\fboxsep=0mm\fbox{\includegraphics[height=5.4cm]{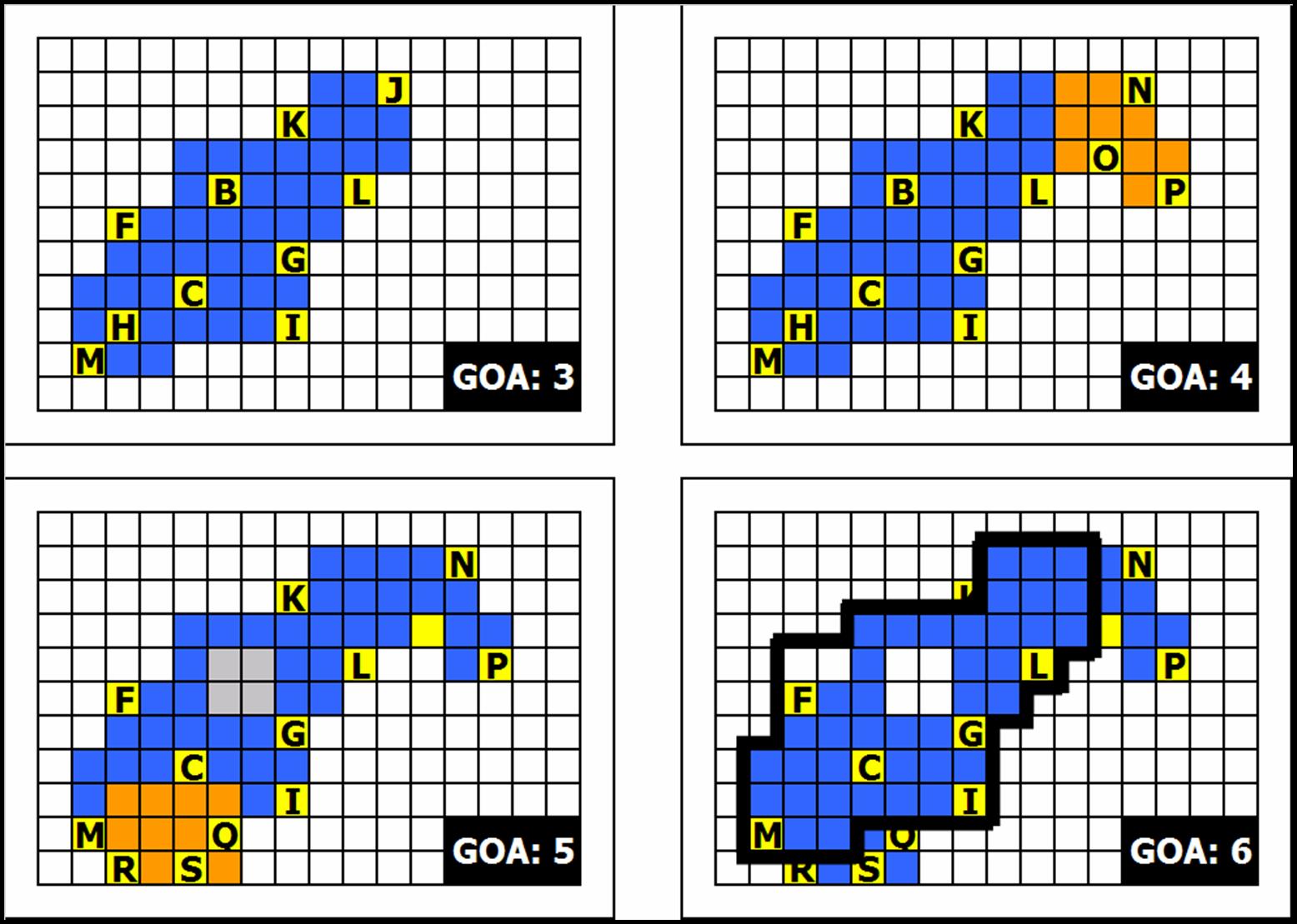}}} 
\caption{Example of artificial ageing. The global clock is allowed to take the values 3-6 after the moment of reproduction. In these steps some operators are activated and affect the phenotype, leading to a worse fitness value in step 6.}
\label{fig13}
\end{center} \end{figure}

\subsection{Artificial Ageing}

As we have seen in previous sections, for a given individual development unfolds in n artificial age steps (n=8 in most experiments performed) and, at the end of it, the individual's fitness value is evaluated. The moment at which the fitness value is evaluated coincides with the moment at which the genome content is frozen and is handed over to the genetic operators of reproduction, cross-over and mutation, in order to be passed on to the next generation. Such moment, that has its biological counterpart in the moment of reproduction, has always coincided in our experiments with the end of the simulation; on the other hand, we can imagine to let the global clock GOA go on and see what happens, in the period after the moment of reproduction. The distinction between the periods before and after reproduction can be thought to correspond to the biological periods of development (say, until 25 years of age in humans) and ageing (from 25 years of age to the moment of death). By analogy, we will call the period (i.e. the set of age steps) {\em before} fitness evaluation ``artificial development'' and the period {\em after} fitness evaluation ``artificial ageing''. 

As pointed out in section 4, at the end of development there are many junk driver cells, as well as many junk operators. This stock of junk represents a reservoir of events that can potentially be triggered after the moment of fitness evaluation, in the period that we called artificial ageing. Since these events occur after fitness evaluation, their effects are by definition not affecting the fitness value; for this reason they will have a random nature: they can be thought of as a random noise superimposed on the phenotype produced by the work of the operators subject to the evolutionary pressure. Given their random nature, the effects of such events on the overall fitness of the phenotype are (much) more likely to be detrimental than beneficial. The set of driver cells not active during development, indicated with DCT-I, can thereforebe further subdivided into a set of driver cells that become active during the ageing period and a set of driver cells that are never active (the true junk): we will refer to these two subsets as to DCT-A (``Ageing'') and DCT-J (``Junk''). Analogously, also the set of operators can be split into a set of operators that become active during the ageing period and a set of operators that are never active, GEN-A and GEN-J respectively. 

\begin{figure}[t!] \begin{center}
{\fboxrule=0.2mm\fboxsep=0mm\fbox{\includegraphics[height=5.4cm]{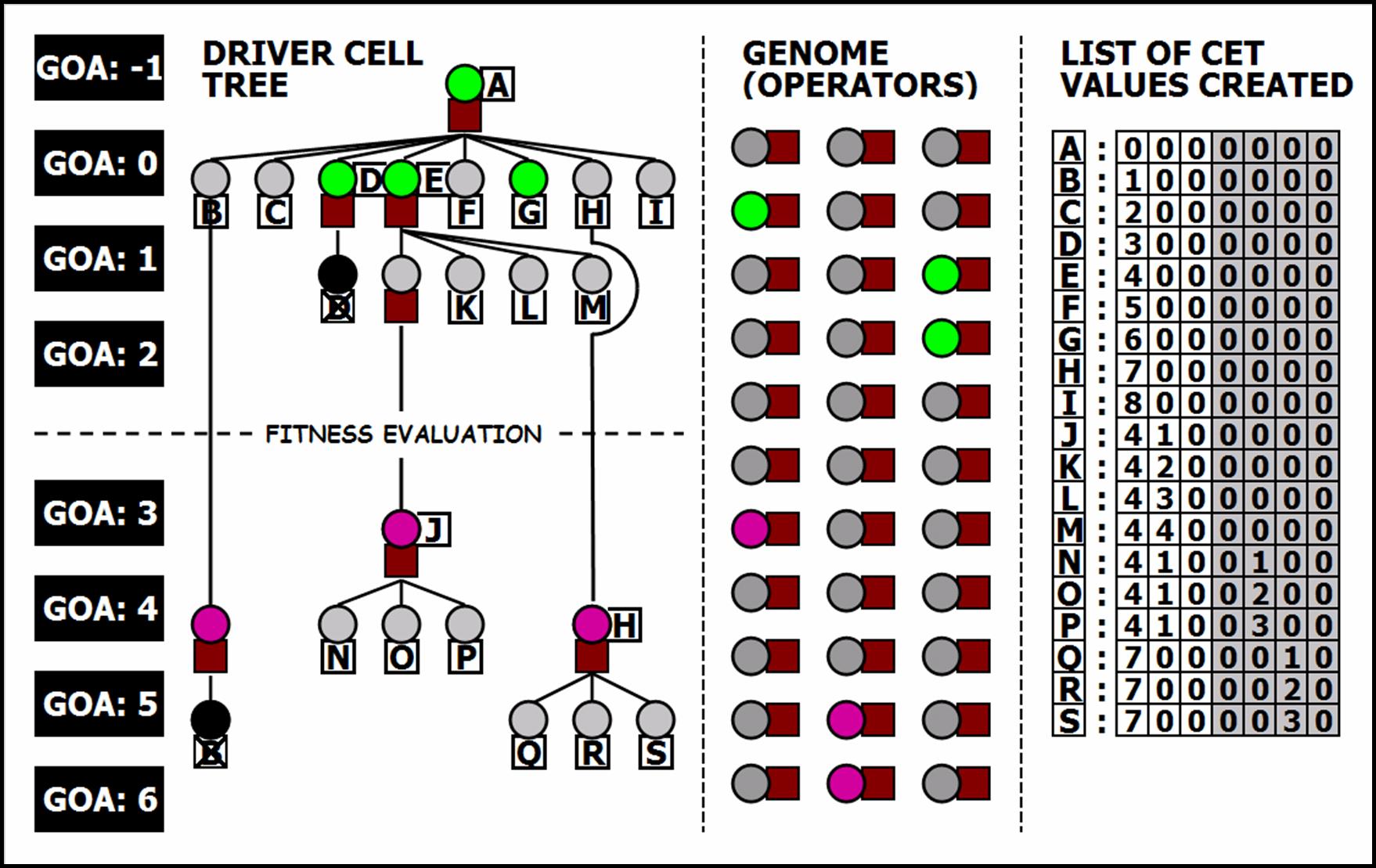}}} 
\caption{Example of artificial ageing. The CET/XET values that are the determinants of the fitness deterioration are shown in purple in both the driver cell tree and the genome. The CET array positions corresponding to the ageing GOA steps are shown in grey.}
\label{fig14}
\end{center} \end{figure}
 
An example of artificial ageing is reported in figure 13 (figure 14 shows the corresponding driver cell tree). The shape evolved in figure 8 is now allowed to develop further (the global clock takes the values 3-6), after the moment of fitness evaluation (which takes place at the end of step 2). In these steps some operators become active and affect the phenotype, leading to a worse fitness value in step 6 (the operators that are the cause of this worsening are shown in purple in the driver cell tree of figure 14). Figure 15 and 16 show a demonstration of this phenomenon for a ``face'' shape (picture of 100x100 size with 16 grey shades); figure 15 refers to the period of development (steps from 0 to 9): the shape grows from the single cell stage to the mature phenotype in step 9, when fitness is evaluated; figure 16 shows steps from 10 to 19, which belong to the period of ageing. As we can see, such period is characterised by the accumulation of random events (both of type proliferation and apoptosis), whose global effect consists in a progressive deterioration of the quality of the image.

\begin{figure}[t] \begin{center}
{\fboxrule=0.2mm\fboxsep=0mm\fbox{\includegraphics[width=8.01cm]{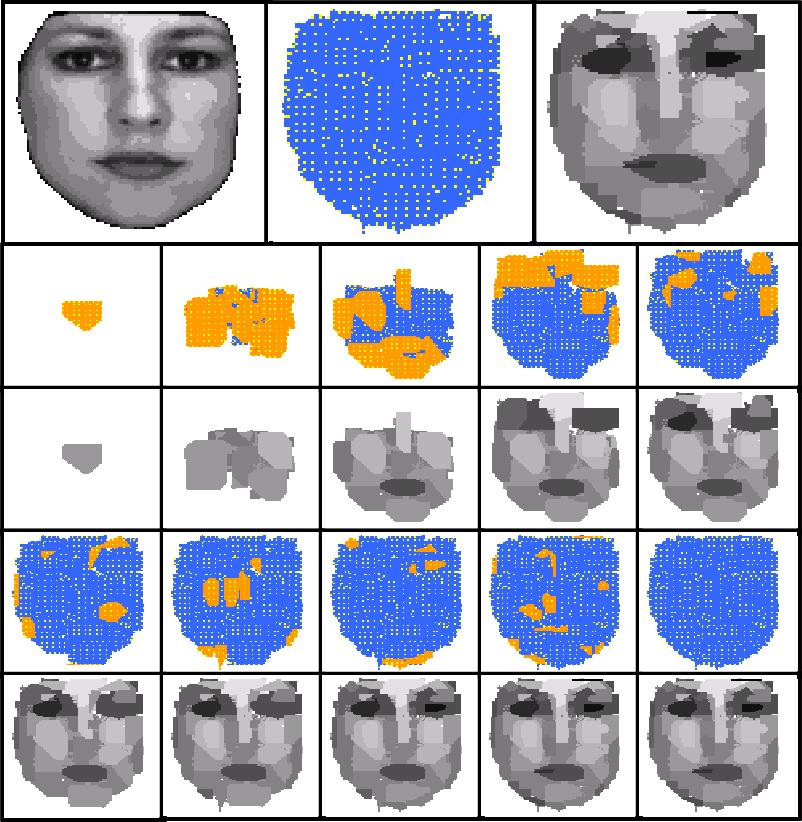}}} 
\caption{The ``face'', period of development (steps 0-9). The shape grows from a single cell to the mature phenotype in step 9, when fitness evaluation takes place}
\label{fig15}
\end{center} \end{figure}

In conclusion, the phenomenon of artificial ageing can be thought as being determined by the cumulative action of change events whose effects manifest themselves after the moment of fitness evaluation, a scenario consistent with the explanation provided by the ``mutation accumulation theory'' for biological ageing. At the end of this section, we wish to dedicate a final remark to the role played by the junk in the phenomenon of artificial ageing. In section 4 the sets of driver cells/operators incative during development, indicated with DCT-I and GEN-I respectively, were shown to be a useful reservoir of new change operators and an indispensable tool to explore new evolutionary paths. In this section we have shown how a part of it (DCT-A and GEN-A) is actually devoted to cause random events that manifest themselves after the moment of reprodution, relegating to DCT-J and GEN-J the role of true junk. On the other hand, we point out how the border between DCT-A/DCT-J and between GEN-A/GEN-J is highly permeable and the average size of DCT-A and GEN-A is proportional to the size of DCT-J and GEN-J (see figure 17). These considerations bring us to deduct, in the epigenetic tracking ``world'', a direct link between the evolvability of a species and its susceptibility to ageing, being both features mediated by the presence of a big stock of junk. The fact that bats have unusually small genomes and display a remarkably long lifespan (i.e. they appear to age less) among mammals of comparable dimension \citep{DY95BL}, could hint to the existence of a similar link also in real biological systems.  

\begin{figure}[t] \begin{center}
{\fboxrule=0.2mm\fboxsep=0mm\fbox{\includegraphics[width=8.01cm]{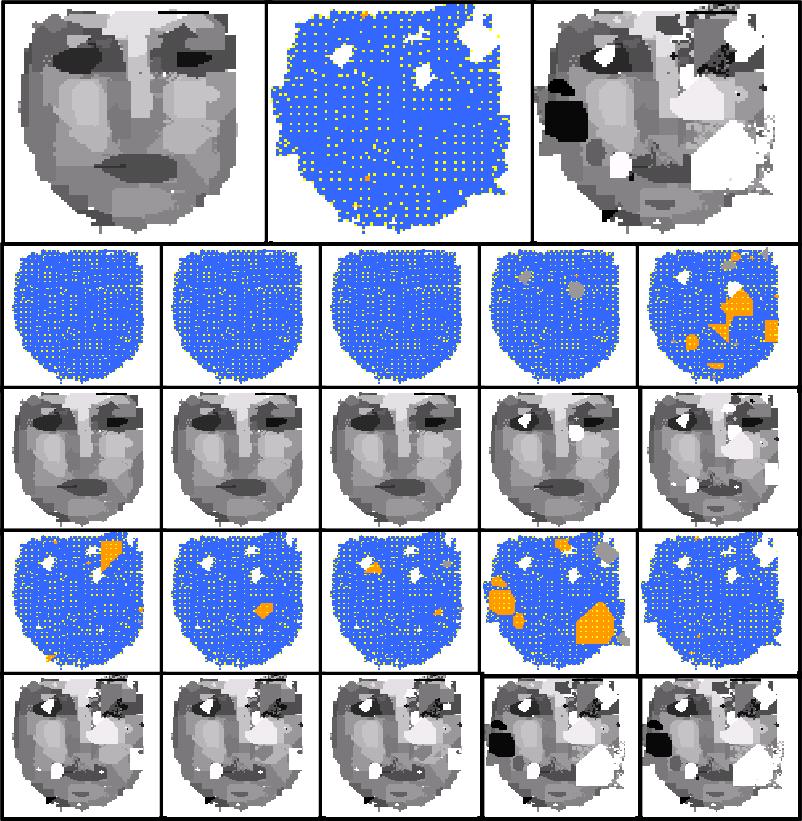}}} 
\caption{The ``face'', period of ageing (steps 10-19). The picture quality deteriores steadily under the action of random operators.}
\label{fig16}
\end{center} \end{figure}

\section{Carcinogenesis}

\subsection{Experimental evidence and theories on carcinogenesis}

\begin{figure}[t] \begin{center}
{\fboxrule=0.2mm\fboxsep=0mm\fbox{\includegraphics[width=8.01cm]{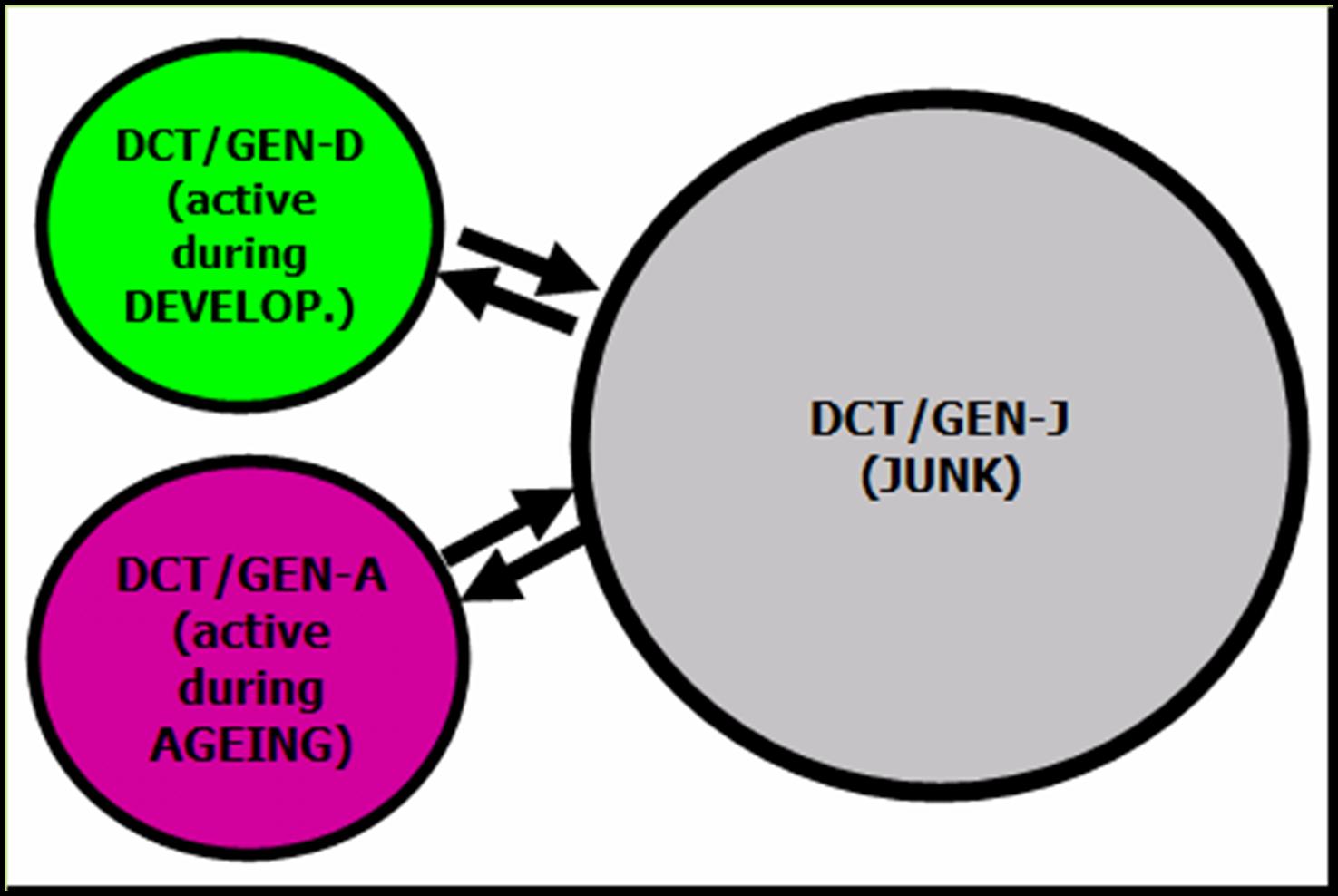}}} 
\caption{The organisation of the DCT and the genome.}
\label{fig17}
\end{center} \end{figure}

So far, we described how the model works under ``normal'', or physiological circumstances. Now we will enter into the realm of artificial pathology; in other words we will analyse possibile malfunctions of the model and we will show how these malfunctions give origin to phenomena that can be considered the artificial equivalent of carcinogenesis, the process by which tumours are formed. Despite having being the subject of intensive investigation, carcinogenesis has failed to disclose all its secrets: as for today, no model has succeeded in the task of explaining all experimental evidence and many questions remain unanswered. In the next sections we will show how, under certain conditions, the model can produce ``malfunctions'' analogous to carcinogenesis; the presence of junk material in both the the driver cell tree and in the genome will once again be shown to play a key role in inducing such malfunctions.

Cancer is a class of diseases in which a group of cells display uncontrolled growth, invasion and sometimes metastasis; these three malignant properties of cancers differentiate them from benign tumors, which are self-limited, do not invade or metastasise. From a cell biology perspective, cancer cells are conjectured to have, compared to normal cells, the following ``superpowers'' \citep{EY00HW}: i) they grow even in the absence of normal ``GO'' signals; ii) they grow despite ``STOP'' commands issued by neighbouring cells; iii) they evade built-in autodestruct mechanisms; iv) they are able to stimulate blood vessel construction; v) they are effectively immortal; vi) they have the power to invade other tissues and spread to other organs. Overall the incidence of cancer rises with age, increasing rapidly during the fourth decade of life (in humans) and continuing to increase thereafter, but more slowly in the fifth, sixth and seventh decades. Epidemiological studies have proved beyond doubt that both genetic and environmental factors are implicated in carcinogenesis. 

The {\bf mainstream theory of carcinogenesis} states that carcinogenesis is a multi-step process that can take place in any cell, driven by damage (mutations) to genes (onco-genes and tumour-suppressor genes) that normally regulate cell proliferation, which in turn upsets the normal balance between cell proliferation and cell death and results in uncontrolled cell division and tumour formation. A more recent theory differentiates from the standard theory in tracing back the origin, the maintenance and the spread of a tumour to a relatively small subpopulation of cells called {\bf cancer stem cells}, whereas the bulk of the tumour would actually be composed of non-tumorigenic cells that, deprived of the cancer stem cells, would quickly shrink and disappear.

A few cancer-related genes, such as p53, do seem to be mutated in the majority of tumors. But many other cancer genes are changed in only a small fraction of cancer types, a minority of patients, or a handful of cells within a tumour. Moreover, some of the most commonly altered cancer genes have oddly inconsistent effects. B. Vogelstein found that the much studied oncogenes c-fosand c-erbb3 are curiously less active in tumors than they are in nearby normal tissues. The tumour suppressor gene RB was recently shown to be hyperactive -not disabled- in some colon cancers, and, perversely, it appears to protect those tumors from their autodestruct mechanisms. In conclusion, the attempt to trace back tumour formation to a subset of mutated genes, consistently found in all tumours, has so far been unsuccessful.

In \citep{EY03GX} three non-standard theories are reviewed. In order to account for the number of mutations (which are very rare events) required to turn a cell cancerous, the {\bf ``modified dogma theory''} \citep{EY03LL} hypothesises that something (a carcinogen, reactive oxidants, or perhaps a malfunction in the cell's DNA duplication and repair machinery) accelerates dramatically the mutation rate; this theory thus adds a prologue to the accepted life history of cancer, but the most important factors are still genetic mutations. In the {\bf ``early instability theory''} \citep{EY01JL}, chromosomal instability occurs early on and represents the first step in carcinogenesis; in this hypothesis, there are several master genes critical for mitosis: if as few as one of these genes is disabled, the cell stumbles each time it divides, muddling some of the chromosomes into an aneuploid state; genetic mutations then lead to a benign tumour, converted later, through additional mutations, to cancer. According to the ({\bf ``all-aneuploidy theory''}, \citep{EY05DX}), cancer cells are aneuploid because they start that way. Many factors can interfere with a dividing cell so that one of its daughter cells is cheated of its normal complement of chromosomes and the other daughter is endowed with a bonus; unlike the other theories, the all-aneuploidy hypothesis predicts that the carcinogenesis is more closely connected to the assortment of chromosomes than to genetic mutations. 

Mathematical models of cancer -see \citep{EY06WK} for a comprehensive review- have found application in three major areas: i) Modelling in the context of epidemiology and other statistical data; ii) Mechanistic modelling of avascular and vascular tumour growth (including physical properties of biological tissues); iii) Modelling of cancer initiation of progression as somatic evolution. Basic mathematical tools used are ordinary differential equations, partial differential equations, stochatic processes, cellular automata and agent-based models. To our knowledge, most mathematical models stick to the standard theory, are based on differential equations and have the primary objective of explaining the dynamics of tumour growth; in such models cells have always been treated as ``black boxes''. Fewer models have tried to open the black box and provide an explanation of carcinogenesis based on the interplay of cell components. 

\subsection{Artificial Carcinogenesis}

In the remainder of this section we will stick to the assumption, shared by most cancer theories, that every tumour has its origin in a single cell, that we will call Originating Cell (OC). The possible occurrences are broken down into four scenarios, characterised by the following driving events:   

\begin{enumerate}
\item Activation of a timed operator in the ageing period (no mutations);
\item Mutation(s) to the cell CET value or to the XET value of an operator in the cell genome, with activation of a junk CET-XET couple;
\item Mutation(s) to the cell CET value or to the XET value of an operator in the cell genome, with activation of a non-junk CET-XET couple;
\item Damage to the mechanism of generation of new CET values.
\end{enumerate}

We will now consider in detail each of these scenarios, analysing the properties of the produced phenomena along three main axes: 1) the rate of incidence in relation to the age; 2) the degree of ``randomness''; 3) the degree of self-sustainability of the growth process. 

\subsection{Scenario 1: activation of a timed operator in the ageing period (no mutations)}

As pointed out in section 5, active driver cells/operators can be subdivided in driver cells/operators active during development and driver cells/operators active during ageing: the first scenario we consider is characterised by the activation of a (CET,XET) couple belonging to (DCT-A,GEN-A), by means of a timed operator. As described in section 2 and recalled in section 4, a timed operator is an operator that is not activated as soon as a matching driver cell appears, but waits until the variable GOA takes a certain predefined value, stored in the operator's variable XOA. In the scenario under consideration, the XOA value is such that GOA takes such value well after development is completed, during what we called the artificial ageing period. If the operator codes for a proliferation event, the cell will proliferate but, unless the CET values generated during such proliferation do not activate other change operators (a very unlikely occurrence), the proliferation will very quickly come to a halt. If the operator codes for an apoptotis event, it will likewise soon be halted (right after its execution).

This scenario is thus characterised by: i) a tendency to occur in the ageing period; ii) a random nature and iii) a non-``self-sustaining'' character and hence a limited duration. The effects on the phenotype can be likened to the superposition of random noise on the concerted ``dance'' of the operators that build the phenotype until the moment of repoduction (and are as such subject to the evolutionary pressure). Given their random nature, the effects of such events on the overall organismal fitness are more likely to be detrimental than beneficial; therefore, we are naturally led to think of the phenomena just described as factors contributing to the process of artificial ageing dealt with in section 5.

A possible biological counterpart of this scenario is represented by a limited, non self-sustaining proliferation of cells, like those giving rise to fibromas (benign tumors composed of fibrous or connective tissue: they can grow in all organs, arising from mesenchyme tissue) or to cysts (closed sacs having a distinct membrane and division on the nearby tissue: they may contain air, fluids, or semi-solid material). These manifestations, even though of proliferative nature, are mostly benign (in that they do not cause a sharp fall in the organismal fitness -only a mild one) and are associated to the ageing phenotype. As may be noted, this scenario does not involve mutations (neither epigenetic nor genetic): malfunctions which are indeed caused by mutations will be examined next.


\subsection{Scenario 2: mutation(s) to the cell CET value (epigenetic) or to the XET value of an operator in the cell genome (genetic), resulting in activation of a junk (CET-XET) couple}

We start by considering the case -scenario 2A- in which the CET of the originating cell suffers an (epigenetic) mutation that turns it into a CET value that activates a change operator in the genetic memory (by matching its XET value). Alternatively -scenario 2B-, it can be the XET of an operator to suffer a (genetic) mutation that turns it into a value that matches the cell's CET value and, as a result, the operator is activated. In a third case -scenario 2C, combination of 2A and 2B- both the CET and the XET are hit by mutations. The common feature of these cases is that a (CET,XET) couple that used to be junk (i.e. used to belong to (DCT-J,GEN-J)), now becomes active. Even though the causes of this scenario (involving epigenetic and/or genetic mutations) and the causes of scenario 1 (no mutations involved) are different, both scenarios have a similar outcome: if the (formerly) junk operator codes for a proliferation event, the cell proliferates but, unless the CET values generated during the proliferation do not activate other change operators (which is very unlikely), the proliferation comes very quickly to a halt (same in case of apoptosis).

Since this scenario is caused by a mutation, in theory it can occur at any age value, either during development or during ageing (the (CET-XET) couple could move from (DCT-J,GEN-J) either to (DCT-D,GEN-D) or to (DCT-A,GEN-A)); on the other hand, since mutations are supposed to be rare events, we can expect it to be more frequent as the age progresses (the tendency of the rate of indidence to increase with the age simply reflecting the time necessary for a relatively rare event to occur); scenario 1, on the other hand, occurs {\em by definition} after the moment of reproduction: for this reason the age-specific incidence rate patterns of scenario 1 and scenario 2, although similar, may not be exactly coincident. This scenario is thus characterised by: i) a tendency to occur more frequently as the age progresses; ii) a ``random noise'' nature and iii) a non-self-sustaining character and hence a limited duration. The biological counterpart is as in scenario 1.


\subsection{Scenario 3: mutation(s) to the cell CET value (epigenetic) or to the XET value of an operator in the cell genome (genetic), resulting in activation of a non-junk CET-XET couple}

\begin{figure}[t!] \begin{center}
{\fboxrule=0.2mm\fboxsep=0mm\fbox{\includegraphics[width=8.01cm]{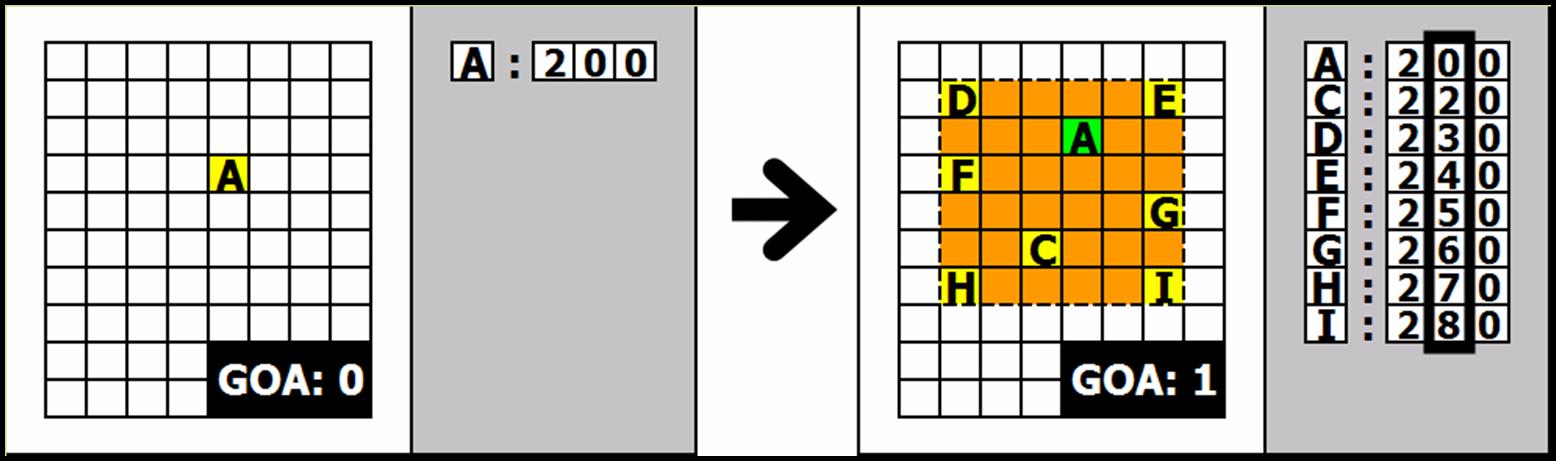}}} 
\caption{``Physiological'' proliferation in maintenance driver cells.}
\label{fig18}
\end{center} \end{figure}

The prologue of scenario 3 is identical to that of scenario 2: a mutation hits either the cell CET variable (scenario 3A) or the XET variable in one of the genome's operators (scenario 3B) or both (scenario 3C), causing the operator to become active and carry out a change event. The key difference is that, while in scenario 2 the activated (CET-XET) couple used to be part of the junk, in scenario 3 it did not. Instead, the CET-XET match and the execution of the relevant event occurred also during (normal) development, where it gave rise to a part of the artificial embryo that contributed to the overall construction of the phenotype: this can be also be expressed by saying that the (CET-XET) couple after the mutation belongs to (DCT-D,GEN-D). Given identical instructions, the cell has no choice but to react identically: therefore it will execute again such part of development, generating other CET values, that in turn will activate other change operators and so forth, until the system comes to a halt, as it did during normal development.

We wish to point out the main characteristic of this scenario. While in the previous scenarios the proliferation comes very quickly to a halt, because the CET values generated are very unlikely to match any change operator in the genome (they are all new, previously unseen values), in the present scenario the first CET value that triggers the change event is indeed a CET value used during development, and so will be (some of) the CET values generated during the proliferation it triggers. In fact this cascade of activations was evolved to construct the phenotype: therefore it will not run out of steam very soon, but will go on for some time, as it did during development. The problem is that now such cascade of activations takes place in the wrong place and in the wrong moment: therefore its overall effect is likely to be akin to a massive perturbation on the evolved phenotype, very different from the random noise typical of the previous scenarios. This scenario is thus characterised by: i) a tendency to occur more frequently as the age progresses (for the rare event effect); ii) a nature non-random noise-like (akin instead to a massive perturbation); iii) a self-sustaining nature, but with a limited duration (as development does not go on forever). 

\begin{figure}[t!] \begin{center}
{\fboxrule=0.2mm\fboxsep=0mm\fbox{\includegraphics[width=8.01cm]{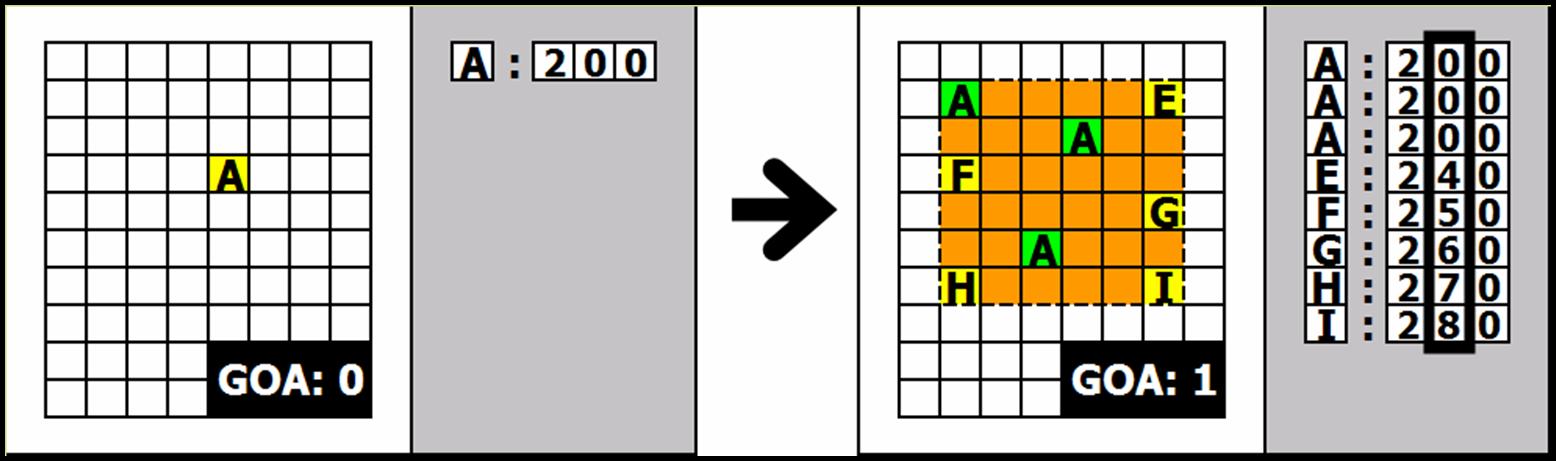}}} 
\caption{Tumorigenic proliferation in maintenance driver cells.}
\label{fig19}
\end{center} \end{figure}

A possible biological counterpart of this scenario is teratoma, a tumour with tissue or organ components resembling normal derivatives of all three germ layers (rarely, not all three germ layers are identifiable). The tissues of a teratoma, although normal in themselves, may be quite different from surrounding tissues, and may be highly inappropriate, even grotesque: teratomas have been reported to contain hair, teeth, bone and very rarely more complex organs such as eyeball, torso, and hand. Usually, however, a teratoma does not contain organs but rather one or more tissues normally found in organs such as the brain, thyroid, liver, and lung. A possible path leading to an (artificial) teratoma is the following: the CET value of an (artificial) adult liver cell is turned into the CET value of a driver cell which in normal development is a precursor of an (artificial) hand; as a result, the mutated driver cell will try to generate the hand. Since such cell finds itself is in the wrong cellular surrounding, it will only manage to minic the normal development of the hand in a grotesque fashion. We finally note that, while in our model the occurrence of the artificial teratoma tends to increase as the age progresses, real teratomas tend to occur more frequently during fetal development. 

  
\subsection{Scenario 4: damage to the mechanism of generation of new CET values during proliferation}

As recalled in section 3, adult stem cells are undifferentiated cells found throughout the body after embryonic development that divide to replenish dying cells and regenerate damaged tissues. In the same section we showed how driver cells can be considered the artificial counterparts of embryonic stem cells; in order to insert also the analogous of adult stem cells in the epigenetic tracking framework we must bring a small modification to the event of proliferation as described in section 2. The modification consists in adding also the CET value ``A'' among the CET values generated during the proliferation event, where ``A'' is the CET value of the mother cell, that triggered the proliferation event in the first place (see figure 18). In other words, the cell with CET value ``A'', while proliferating, produces a copy of itself, that in turn will give rise to another (identical) proliferation event.

In physiological conditions we can think that the rate at which ``A'' driver cells are generated is exactly counterbalanced by the rate at which ``A'' driver cells are consumed or destroyed, in such a way that the net balance is zero (system in equilibrium). If we call ``percentage of replenishment'' (indicated with P) the percentage of ``A'' driver cells on the total of daughter cells, we can refer to P0 as the value that keeps the system in equilibrium. This functioning corresponds to the behaviour of adult stem cells, that are generated to replenish the stem pool whenever it becomes depleted (because too many stem cells have gone down a differentiation path to perform their specialised jobs in the organism). To distinguish the newly introduced kind of proliferation from the standard one, we will call the events/driver cells/operators active during development ``{\bf development} events/driver cells/operators'', while those acting after development (i.e. during the ageing period, equivalent to adult stem cells) will be called ``{\bf maintenance} events/driver cells/operators''.   

Now, the stage for a more dangerous scenario (scenario 4A) is set when a fault arises in the mechanism that cells use to generate new CET values during a maintenance event of proliferation type. Within this scenario many variants are conceivable (such mechanism can be damaged in many ways): in one possible variant the damage can be such that the percentage P is increased beyond the value P0 that maintains the system in equilibrium. Figure 19 shows an example: instead of having only one cell bearing the same CET value of the mother, now there are three. In this situation the originating cell and its epigenetically identical progeny are stuck to execute the same operator, leading to an increased proliferation rate that overshoots the destruction rate; the ultimate effect is what will be perceived as uncontrolled proliferation: the distinguishing mark of tumours. Scenario 4A is thus characterised by: i) a tendency to occur later in life; ii) a nature non-random noise-like; iii) a self-sustaining nature, with {\em unlimited} duration. 


In the scenario just described the originating cell is a maintenance driver cell, in which even under normal circumstances a fixed percentage of the progeny shares the same CET value of the mother. But the same thing could also take place in a development driver cell, in which under normal circumstances no driver cells in the progeny share the CET value of the mother. In this scenario (4B) cellular proliferation behaviour goes from the paradigm of figure 1 directly to the paradigm of figure 19. The damage to the CET value mechanism generation can occur either in driver cell active during development or in a driver cell active during ageing; in this latter case, since the operator involved is always timed, a peculiar phenomenon takes place: even though the damage occurs early in life (in theory it can occur at any artificial age), in order to manifest itself the damage must wait until GOA takes the value stored in the operator's XOA. This means that, between the moment of the damage and the moment at which its effects become manifest, there can be a long latency period. Scenario 4B is thus characterised by: i) a tendency to occur later in life; ii) a nature non-random noise-like; iii) a self-sustaining growth pattern, with {\em unlimited} duration.     


The biological counterpart for both scenario 4A and 4B is represented by benign tumours, expanding with varying degrees of speed, which roughly corresponds to ``superpowers'' 1 and 2 of subsection 6.1: further genetic mutations are hypothesised to be necessary in order to acquire the additional superpowers that are the hallmark of cancer. In this view carcinogenesis is driven by a combination of structural damage (the biological equivalent of the CET value generation mechanism) and genetic mutations; this appears to fit quite well with the experimental evidence, that seems to require also a structural damage in the cell, in additions to genetic mutations, as a requirement for carcinogenesis (requirement acknowledged in the ``early instability theory'' and in the ``all-aneuploidy theory''). The long latency between the supposed moment at which the damage occurs (e.g. exposure to tobacco smoke) and the moment at which cancer breaks out (e.g. lung cancer) is a well known phenomenon, whose artificial counterpart, as we have seen, our model is able to reproduce. In conclusion, our model can be seen as a combination of the ``cancer stem cell theory'' and the ``all-aneuploidy theory''. 

\section{Conclusions}

We presented a model of cellular growth called ``epigenetic tracking'' and demonstrated its effectiveness in generating arbitrary shapes by means of evo-devo techniques. The model was subsequently applied to the study of key biological issues such as embryogenesis, the organisation of genome, ageing and carcinogenesis, showing that is can produce artificial counterparts for each of them. We can conclude that the model proposed has the potential to be the foundation for a project to build a whole artificial biology, displaying many aspects in common with real biology. Future work will be focused on some key issues, that at present are not biologically plausible, namely i) modelling of inter-cellular signalling and ii) modelling the genetic regulatory network that is known to function in real cells.


\begin{thebibliography}{}

\bibitem[Ahnert and et~al., 2008]{CY08AX}
Ahnert, S.~E. and et~al. (2008).
\newblock How much non-coding dna do eukaryotes require?
\newblock {\em J Theor. Biol.}, 252:587--592.

\bibitem[Bongard and Pfeifer, 2001]{AY01BP}
Bongard, G. and Pfeifer, R. (2001).
\newblock Repeated structure and dissociation of genotypic and phenotypic.
  complexity in artificial ontogeny.
\newblock In {\em Proceedings of The Genetic and Evolutionary Computation
  Conference, GECCO-2001}, pages 829--836. Morgan Kaufmann.

\bibitem[Cangelosi et~al., 1994]{AY94CN}
Cangelosi, A., Nolfi, S., and Parisi, D. (1994).
\newblock A gene network model for developing cell lineages.
\newblock {\em Artificial Life}, 5:497--515.

\bibitem[De~Garis, 1999]{AY99DG}
De~Garis, H. (1999).
\newblock {\em Artificial Embryology and Cellular Differentiation}.
\newblock Academic Press.

\bibitem[Deininger and Batzer, 2002]{CY02DB}
Deininger, P.~L. and Batzer, M.~A. (2002).
\newblock Mammalian retroelements.
\newblock {\em Genome Res.}, 12 (10):1455--1465.

\bibitem[Duesberg, 2005]{EY05DX}
Duesberg, P. (2005).
\newblock Does aneuploidy or mutation start cancer?
\newblock {\em Science}, 307(5706):41.

\bibitem[Fontana, 2008]{AY08AX}
Fontana, A. (2008).
\newblock Epigenetic tracking, a method to generate arbitrary shapes by using
  evo-devo techniques.
\newblock In {\em Epirob08}.

\bibitem[Gibbs, 2003]{EY03GX}
Gibbs, W.~W. (2003).
\newblock Untangling the roots of cancer.
\newblock {\em Scientific American}, 289:56--65.

\bibitem[Gruau et~al., 1996]{AY96GW}
Gruau, F., Whitley, D., and Pyeatt, L. (1996).
\newblock A comparison between cellular encoding and direct encoding for
  genetic neural networks.
\newblock In {\em Genetic Programming 1996}, page xxxx.

\bibitem[Hanahan and Weinberg, 2000]{EY00HW}
Hanahan, D. and Weinberg, R. (2000).
\newblock The hallmarks of cancer.
\newblock {\em Cell}, 100:57--70.

\bibitem[Hogeweg, 2003]{AY03HX}
Hogeweg, P. (2003).
\newblock Computing an organism: on the interface between informatic and
  dynamic processes.

\bibitem[Hornby and Pollack, 2002]{AY02HP}
Hornby, G.~S. and Pollack, J.~B. (2002).
\newblock Creating high-level components with a generative representation for
  body-brain evolution.
\newblock {\em Artificial Life}, 8(3).

\bibitem[Jallepalli and Lengauer, 2001]{EY01JL}
Jallepalli, P.~V. and Lengauer, C. (2001).
\newblock Chromosome segregation and cancer: Cutting through the mystery.
\newblock {\em Nature Reviews Cancer}, 1:109--117.

\bibitem[Kauffman, 1969]{AY69KF}
Kauffman, S.~A. (1969).
\newblock Metabolic stability and epigenesis in randomly constructed genetic
  nets.
\newblock {\em Journal of Theoretical Biology}, 22:437--467.

\bibitem[Kirkwood, 1977]{DY77KX}
Kirkwood, T. (1977).
\newblock Evolution of ageing.
\newblock {\em Nature}, 270:301--304.

\bibitem[Kumar and Bentley, 2003]{AY03KB}
Kumar, S. and Bentley, P.~J. (2003).
\newblock {\em On Growth, Form and Computers}.
\newblock Academic Press.

\bibitem[Lindenmayer, 1968]{AY68LX}
Lindenmayer, A. (1968).
\newblock Mathematical models for cellular interaction in development.
\newblock {\em Journal of Theoretical Biology}, 18:280--289.

\bibitem[Loeb et~al., 2003]{EY03LL}
Loeb, L.~A., Loeb, K.~R., and Anderson, J.~P. (2003).
\newblock Multiple mutations and cancer.
\newblock {\em Proceedings of the National Academy of Sciences USA},
  100-3:776--781.

\bibitem[Miller and Banzhaf, 2003]{AY03MB}
Miller, J.~F. and Banzhaf, W. (2003).
\newblock Evolving the program for a cell: from french flags to boolean
  circuits.
\newblock In Press, A., (Ed.), {\em On Growth, Form and Computers}.

\bibitem[Piegu and et~al., 2006]{CY06PX}
Piegu, B. and et~al. (2006).
\newblock Doubling genome size without polyploidization: Dynamics of
  retrotransposition-driven genomic expansions in oryza australiensis, a wild
  relative of rice.
\newblock {\em Genome Res.}, 16:1262--1269.

\bibitem[Sims, 1994]{AY94SX}
Sims, K. (1994).
\newblock Evolving 3d morphology and behavior by competition.
\newblock In Brooks, R.~A. and (Eds.), P.~M., (Eds.), {\em Proceedings of
  Artificial Life IV}, pages 28--39.

\bibitem[Stanley and Miikkulainen, 2003]{AY03SM}
Stanley, K. and Miikkulainen, R. (2003).
\newblock A taxonomy for artificial embryogeny.
\newblock {\em Artificial Life}, v.9 n.2:93--130.

\bibitem[Turing, 1952]{AY52TX}
Turing, A. (1952).
\newblock The chemical basis of morphogenesis.
\newblock {\em Philosophical Transactions of the Royal Society}, 237:37--72.

\bibitem[Van~den Bussche et~al., 1995]{DY95BL}
Van~den Bussche, R.~A., Longmire, J.~L., and Baker, R.~J. (1995).
\newblock How bats achieve a small c-value: frequency of repetitive dna in
  macrotus.
\newblock {\em Mamm Genome}, 6(8):521--525.

\bibitem[Wodarz and Komarova, 2006]{EY06WK}
Wodarz, D. and Komarova, N.~L. (2006).
\newblock {\em Computational biology of cancer}.
\newblock World Scientific.

\bibitem[Woolfe, 2005]{CY05WX}
Woolfe, A., e.~a. (2005).
\newblock Highly conserved non-coding sequences are associated with vertebrate
  development.
\newblock {\em PLoS Biol}.

\end{thebibliography}
\end{document}